\documentclass[fleqn,usenatbib]{mnras}

\usepackage[T1]{fontenc}
\usepackage{ae,aecompl}

\usepackage{graphicx}	
\usepackage{subfig}        
\usepackage{amsmath}	
\usepackage{amssymb}	

\usepackage{lscape}     

\def\Lya{Ly$\alpha$}

\def\HI{\hbox{H~$\scriptstyle\rm I\ $}}

\def\HII{\hbox{H~$\scriptstyle\rm II\ $}}

\def\kms{\,{\rm {km\, s^{-1}}}}
\def\msun{{M_\odot}}

\def\ltsima{$\; \buildrel < \over \sim \;$}
\def\lsim{\lower.5ex\hbox{\ltsima}}
\def\gtsima{$\; \buildrel > \over \sim \;$}
\def\gsim{\lower.5ex\hbox{\gtsima}}

\title{The effect of ionizing background fluctuations on the spatial correlations of high redshift Ly$\alpha$-emitting galaxies}

\author[A. Meiksin \& T. Suarez]{Avery Meiksin$^{1}$ \thanks{Contact e-mail: \href{mailto:aam@roe.ac.uk}{meiksin@ed.ac.uk}}
\& Teresita Suarez $^{1}$
\\
$^{1}$SUPA\thanks{Scottish Universities Physics Alliance}, The Royal Observatory, Edinburgh, Blackford Hill, Edinburgh EH9 3HJ, UK
}

\date{Accepted 8 August 2022. Received 5 August 2022; in original form 30 May 2022.}

\pubyear{2022}
\begin{document}
\label{firstpage}
\pagerange{\pageref{firstpage}--\pageref{lastpage}}
\maketitle

\begin{abstract}
  We investigate the possible influence of fluctuations in the
  metagalactic photoionizing ultra-violet background (UVBG) on the
  clustering of \Lya-emitting galaxies through the modulation of the
  ionization level of the gas surrounding the systems. At redshifts
  $z>5$, even when assuming the reionization of the intergalactic
  medium has completed, the fluctuations are sufficiently large that
  they may non-negligibly enhance, and possibly even dominate, the
  angular correlation function on scales up to a few hundred
  arcsecs. Whilst a comparison with observations at $z\simeq5.7$ is
  statistically consistent with no influence of UVBG fluctuations,
  allowing for the fluctuations opens up the range of acceptable
  models to include those with relatively low bias factors for the
  \Lya-emitting galaxies. In this case, the evolution in the bias
  factor of \Lya-emitters over the approximate redshift range $3<z<7$
  corresponds to a nearly constant halo mass for \Lya-emitting
  galaxies of $\sim10^{10.5}\msun$.
  \end{abstract}

\section{Introduction}
\label{sec:Intro}

The discovery of high redshift \Lya-emitting galaxies
\citep{1996Natur.382..231H, 1996Natur.383...45P} has led to a series
of \Lya\ emitter (LAE) surveys of increasing depth
\citep{2020ARA&A..58..617O}. Surveys at $z>7$ likely probe the Epoch
of Reionisation (EoR), when the Universe was transformed from neutral
to ionized by the first galaxies and Quasi-Stellar Objects (QSOs),
over a characteristic redshift range $6\lsim z\lsim10$
\citep{2020A&A...641A...6P}. LAE surveys therefore provide not only
information about some of the properties of the earliest galaxies in
the Universe, but also about the evolving nature of the Intergalactic
Medium (IGM).

The evolution in the luminosity distribution of LAEs has been used to
constrain the ionization state of the IGM during the EoR. The decrease
in the LAE \Lya\ luminosity function compared with the ultra-violet
(UV) continuum luminosity function of Lyman-break galaxies (LBGs)
agrees with an increasing neutral hydrogen fraction of the IGM with
redshift. The mean neutral hydrogen fraction is estimated to be
$\bar x_\textrm{HI}>0.1$ at $z>6.6$ \citep{2018PASJ...70S..16K}, and
it possibly exceeds $\bar x_\textrm{HI}>0.2$ at $z>7$
\citep[eg][]{2018ApJ...867...46I, 2021ApJ...923..229G}. The evolving
rate of detection of \Lya\ emission in LBGs was used by
\citet{2018ApJ...856....2M} to suggest
$\bar x_\textrm{HI}=0.59^{+0.11}_{-0.15}$ at $z\sim7$.

The effects of incomplete reionization on the properties of LAEs are
less able to constrain the ionization state of the IGM at $z<6$. The
strongest constraints come from the \Lya\ forest measured in bright,
background QSOs. The distributions of the measured \Lya\ effective
optical depths $\tau_\alpha$ at $z>5$ are unexpectedly broad compared
with the predictions of numerical simulations of the IGM. The breadths
of the distributions increase with redshift, with patches of high
attenuation extending over 100~$h^{-1}$ comoving Mpc appearing along
some lines of sight at $z>6$ \citep{2015MNRAS.447.3402B,
  2018ApJ...863...92B, 2018MNRAS.479.1055B, 2020ApJ...904...26Y,
  2022MNRAS.514...55B}. It has been suggested that late reionization
may account for the breadths of the \Lya\ effective optical depth
distributions, with the EoR ending as late as $z\simeq5.2-5.5$
\citep{2019MNRAS.485L..24K, 2020MNRAS.491.1736K,
  2020MNRAS.494.3080N}. In this scenario, extended regions of high
optical depth originate in residual patches of neutral hydrogen.

An alternative explanation suggested for the breadths of the
distributions is the \lq\lq dappling\rq\rq\ of the IGM by a
fluctuating metagalactic UVBG
\citep{2020MNRAS.491.4884M}. Fluctuations in the UVBG arise from both
spatial correlations between the sources and from the shot noise due
to their discreteness \citep{2014MNRAS.442..187G, 2014PhRvD..89h3010P,
  2017MNRAS.472.2643S, 2019MNRAS.482.4777M}. Whilst the fractional
contribution of QSO sources to the UV background is below half at
$z>3$ and diminishes with increasing redshift, the QSOs continue to
dominate the shot noise contribution to the background. The large
scale UVBG fluctuations resulting from the source shot noise reproduce
the measured breadths in the distributions of the \Lya\ effective
opacity at $3<z<6$.

LAEs have been measured to cluster spatially. The strength of the
  clustering has been used to infer the nature of LAEs from the
  estimated bias factors. Comparison with the expected clustering of
dark matter shows an increase in the LAE bias from $b_L\sim1.5$ at
$z<4$ to as high as $b_L\sim6$ at $z>5$, with inferred dark matter
halo masses of $10^{10.3}-10^{11.3}\,{\rm M}_\odot$ and a duty cycle,
representing the fraction of haloes in a LAE phase, of less than 1\%
\citep{2018PASJ...70S..13O}. From the evolution of the bias factor,
\citet{2018PASJ...70S..13O} infer LAEs at z > 2 may be the progenitors
of massive elliptical galaxies today. The estimates of the LAE bias
factors, however, show considerable scatter at the higher redshifts,
with estimates for $b_L$ at $z=5.7$ ranging between $4.1\pm0.2$ and
$6.1\pm0.7$, and at $z=6.6$ between $b_L=3.6\pm0.7$ and $4.5\pm0.6$,
depending on the field(s) analysed \citep{2010ApJ...723..869O,
  2018PASJ...70S..13O}. The origin of the discrepant values,
especially at $z=5.7$, is unclear. It has been suggested that results
for some of the smaller fields may be affected by cosmic variance
\citep{2018PASJ...70....4K}, with the clustering in one field at
$z=5.7$ dominated by a protocluster
\citep{2010ApJ...723..869O}. Firmly establishing the evolution of the
LAE bias factors will help elucidate the nature of the host systems at
high redshifts and of their counterparts today.

The interpretation of the clustering signal is also complicated
  by the possible effects of \Lya\ photon scattering by the
  intervening IGM. The clustering of LAEs at $z>5$ may depend on the
  physical character of the EoR through the ionization structure of
  the IGM. The patchwork of ionized and still-neutral regions during
the EoR will be imprinted on the spatial correlations of LAEs because
of the relative difference in attenuation of the Ly$\alpha$ emission
line of the galaxies in different patches \citep{2006MNRAS.365.1012F,
  2007MNRAS.381...75M}. Comparison with observations suggests a mean
IGM neutral hydrogen fraction below 30\% at $z=6.6$
\citep{2018PASJ...70S..13O}.

The clustering of LAEs may also depend on radiative transfer
  effects on the sample selection, since the \Lya\ emission line may
  be modulated by scattering through the intervening IGM, which traces
  the same dark matter density field as do the galaxies
\citep{2011MNRAS.415.3929W, 2011ApJ...726...38Z,
  2020MNRAS.491.3266G}. Local non-linear redshift space
  distortions may further confound interpretation of the clustering
  strength \citep{2019MNRAS.489.3472B}.

In this paper, we suggest that another factor governing the
clustering of LAE systems are fluctuations in the UVBG after the
end of cosmic reionization. The gas surrounding a LAE may give rise to
a non-negligible \Lya\ optical depth redward of the \Lya\ emission
line \citep[eg][]{2006A&A...460..397V, 2011ApJ...728...52L}. UVBG
fluctuations will then modify the detectability of the systems and
imprint their spatial correlations on the spatial correlation function
of the LAEs. We assess the possible contribution of UVBG fluctuations
to the spatial correlations of LAEs, and constrain the attenuating
properties of the gas surrounding the LAEs. We shall show that
allowing for UVBG fluctuations increases the uncertainty in the LAE
bias factors inferred from the angular correlation function.

Modelling the effect of the UVBG on the spatial correlations of LAEs,
however, is hampered by two difficulties:\ (1)\ the rarity of the QSOs
at high redshift requires either very large simulation volumes, on the
order of 800$h^{-1}$~Mpc (comoving) or very many statistical
realisations in smaller volumes, to capture the shot-noise
contribution \citep{2020MNRAS.491.4884M}, and (2)\ the origin and
subsequent radiative transfer of the \Lya\ emission line from LAEs is
unknown. We overcome the first difficulty by using a perturbative
approach to model the UVBG fluctuations \citep{2019MNRAS.482.4777M},
as the underlying density fluctuations are small over the physical
scales for which we estimate the LAE spatial correlations. The latter
difficulty includes uncertainties in the amount of obscuration of the
escaping \Lya\ photons from the unknown clumpiness and flow pattern of
the surrounding circumgalactic and intergalactic media. We adopt a
simple statistical model described below to assess the amount of
obscuration of the \Lya\ emission line by the gas surrounding
LAEs. The combined model is designed to capture the range of possible
impact the UVBG fluctuations may have on the LAE spatial correlations.

The paper is organised as follows. In the next section, we describe
our model for assessing the possible effect of UVBG fluctuations on
the measured angular correlation function of LAEs. In Sec.~3,
comparisons between the model predictions and measurements are
presented. We discuss these in Sec.~4, and provide a summary of our
conclusions in the last section. A flat cosmology is assumed
  with $\Omega_m=0.31$, $\sigma_8=0.81$ and $n=0.97$
  \citep{2020A&A...641A...6P}.

\section{The model}
\label{sec:model}

\subsection{UV background model}
\label{subsec:uvbg}

The mean photoionizing UVBG is modelled following
\citet{2019MNRAS.482.4777M} and \citet{2020MNRAS.491.4884M}, with some
recent updates. In brief, both galactic and QSO sources are included
in the mean metagalactic emissivity. The evolving Schechter luminosity
function fit from \citet{2015ApJ...803...34B} is adopted for the
galactic contribution. We use Model 3 of \citet{2019MNRAS.488.1035K}
for the QSO luminosity function, confined to sources with absolute AB
magnitudes at 1450~A of $-30<M_{1450}<-21$. The mean free path of
ionizing photons at the Lyman photoelectric edge is taken from
\citet{2014MNRAS.445.1745W} for $z<5$ and \citet{2021MNRAS.508.1853B}
for $z\ge5$. The mean metagalactic emissivity at the Lyman edge is
adjusted to reproduce the median effective IGM \Lya\ optical depth
measurements listed in \citet{2009RvMP...81.1405M} for $z<5$, and from
\citet{2018MNRAS.479.1055B} for $5\leq z<5.4$ and
\citet{2020ApJ...904...26Y} for $z\geq5.4$.

The fluctuations in the UVBG arise from fluctuations in the Lyman
continuum opacity of the IGM, and from the spatial clustering of the
sources and the source shot noise. We assume a galaxy bias factor
$b_G=3$ \citep[e.g.][]{2013MNRAS.430..425B}, noting it may be higher
at $z>5$. An evolving bias factor is adopted for the QSOs of
$b_Q=0.278(1+z)^2+0.57$ \citep{2017JCAP...07..017L}. At $z>5$, the
fluctuations are dominated by the shot noise in the QSO counts over
the scales of interest. For a short source lifetime compared with the
age of the Universe, the shot noise contribution is proportional to
the source lifetime. For a long lifetime, the shot noise is given by
the steady-state limit.

Following \citet{2019MNRAS.482.4777M}, we define the following
quantities:\ for an UVBG intensity $I_\nu$, emissivity $j_\nu$ and
photoelectric cross section $\sigma_\nu$,
$f=\int\,d\nu (I_\nu/h_\mathrm{P}\nu)\sigma_\nu$,
$j=\int\,d\nu (j_\nu/h_\mathrm{P}\nu)\sigma_\nu$ and
$\zeta=\langle I_L\rangle\sigma_L/h_\mathrm{P}\langle f\rangle$, where
$h_\mathrm{P}$ is Planck's constant and $I_L$ and $\sigma_L$ are the
values of the intensity and cross section at the Lyman edge,
respectively. Here, $\langle \dots\rangle$ denotes a spatial
average. A dimensionless opacity
$\chi=(c/H)\langle\alpha_\mathrm{eff}\rangle$ is also defined, where
$\alpha_\mathrm{eff}$ is the frequency-averaged opacity weighted by
$I_\nu\sigma_\nu/h_\mathrm{P}\nu$, and $H$ is the Hubble
parameter. The dimensionless ratio
$\phi=c\langle j\rangle/[H(\chi+\zeta)\langle f\rangle]$ describes the
evolution of the radiation field. For a non-evolving radiation field,
$\phi=1$. The mean metagalactic photoionization rate is related to $f$
by $\Gamma=4\pi\langle f\rangle$. The Fourier component for comoving
wavenumber $k$ of the perturbation to the photoionization rate is
given in the steady-state limit by
\begin{equation}
  \tilde\delta_{\Gamma, {\rm SS}}(\kappa)=\frac{\phi(\chi+\zeta)\tilde\delta_j(\kappa)
    - b_{\chi, \delta}\chi\tilde\delta(\kappa)}{{\frac{\kappa}{a}\left[{\rm
    atan}{\left(\frac{\kappa}{a\phi(\chi+\zeta)}\right)}\right]^{-1}}
+b_{\chi, \Gamma}\chi},
\label{eq:dGisofullSS}
\end{equation}
where $\tilde\delta_j$ and $\tilde\delta$ are the perturbations in the
source density and gas density respectively, $\kappa=(c/H)k$ is a
dimensionless comoving wavenumber, and $a=1/(1+z)$. Here,
$\tilde\delta\log\chi =
b_{\chi,\delta}\tilde\delta+b_{\chi,\Gamma}\tilde\delta_\Gamma$. Eq.~(\ref{eq:dGisofullSS})
is similar to the findings of \citet{2014MNRAS.442..187G} and
\citet{2014PhRvD..89h3010P}, who both additionally assume
$\phi=1$. The power spectrum of the light fluctuations in a comoving
volume $V_u$ is
$P_\Gamma^\mathrm{SS}(k)=V_u\langle
\tilde\delta_\Gamma^\dagger(k)\tilde\delta_\Gamma(k)\rangle^\mathrm{SS}$. In
addition to a component proportional to the dark matter power spectrum
and dependent on the bias factors of the sources, a source shot noise
term also contributes additively to the full power spectrum. The shot
noise contribution is given by
\begin{equation}
V_u\langle
\tilde\delta^*_\Gamma(k)\tilde\delta_\Gamma(k)\rangle^\mathrm{SS}_\mathrm{shot}\sim\left[\frac{\pi}{2}\phi\left(\chi+\zeta\right)\frac{a}{\kappa}\right]^2\frac{1}{n_{\rm eff}},
\label{eq:shothighk}
\end{equation}
in the limit $\kappa\gg1$, corresponding to modes with wavelengths
short compared with the cosmological horizon. The effective number
density of the sources is given by $n_\mathrm{eff}=\langle
L\rangle_\Phi^2/\langle L^2\rangle_\Phi$, where the averages
$\langle\dots\rangle_\Phi$ are carried out over the luminosity
density $\Phi(L)$ of the sources.

For sources with finite lifetimes $\tau_S$, the shot noise contribution may be
substantially reduced. We adopt an approximate form which accurately
interpolates between the $\kappa\ll1$ and $\kappa\gg1$ limits,
\begin{equation}
V_u\langle
\tilde\delta^*_\Gamma(k)\tilde\delta_\Gamma\rangle_{\rm shot}(k)\simeq
\frac{1}{A}\frac{\left[\phi(\chi+\zeta)\right]^2}{1 + [\kappa/\kappa_*^{\rm SN}(\kappa)]^2}\frac{1}{n_{\rm eff}},
\label{eq:apshotLor}
\end{equation}
where
\begin{eqnarray}
  A &=& \frac{2}{H\tau_S}\left[(\phi+b_{\chi, \Gamma})\chi+\phi\zeta+\frac{3}{4}-\frac{1}{2}\alpha_n\right]\\
  &&-\left[\frac{4}{\pi^2}\left(\phi(\chi+\zeta)-1/2\right)+b_{\chi, \Gamma}\chi\right]^2,\nonumber
  \label{eq:defAlor}
\end{eqnarray}
  for the approximation $n_\mathrm{eff}\sim(1+z)^{-\alpha_n}$, and $\kappa_*^{\rm SN}(\kappa) = 2\pi[c/H(z)]/[(1+z) \lambda_*^{\rm SN}(\kappa)]$ with
\begin{eqnarray}
  \lambda_*^{\rm SN}&=&4\frac{c}{H}A^{-1/2}\\
  &&\times\left\{1+\frac{\pi}{2}\left[\frac{4}{\pi^2}\left(\phi(\chi+\zeta)-1/2\right)+b_{\chi, \Gamma}\chi\right]\frac{a}{\kappa(a)}\right\}\nonumber,
  \label{eq:lambdas_shot}
\end{eqnarray}
for $H\tau_S\ll1$ \citep{2019MNRAS.482.4777M}. When the distance light
travels during the source lifetime exceeds the mean free path of the
ionizing radiation, the steady-state value is a good approximation. On
scales small compared with the light travel distance over the source
lifetime, $k\gg 1/(c\tau_S)$, Eq.~(\ref{eq:apshotLor}) goes over to
Eq.~(\ref{eq:shothighk}). Additional suppression of the power spectrum
of UVBG fluctuations may arise if QSOs emit in tight beams
\citep{2017MNRAS.472.2643S}; this effect is not included here.

When including the UVBG fluctuations, we adjust the escape fraction of
ionizing photons from galaxies to match the median optical depth
measurements. In practice this has only a small effect on the shot
noise contribution because it is dominated by the QSOs.

\subsection{Observed \Lya\ emission line}
\label{subsec:lyaline}

Modelling the \Lya\ emission line is more problematic. The observed
line depends on the emission line mechanism and the subsequent \Lya\
radiative transfer both within and near the \Lya-emitting galaxy and
through the IGM. The emission line may originate in a compact region
or from several regions extending across the galaxy, which will affect
the line profile generated. This will also affect the \Lya\ radiative
transfer through the interstellar medium of the galaxy, which may be
static or moving, possibly in a wind. The profile will be further
modified by the gas in the circumgalactic medium (CGM) of the galaxy,
and whether the gas is expanding in a wind or infalling, and possibly
part of an \HII\ region produced by the LAE. The possible presence of
dust absorption is still another complicating factor. All of these
factors may vary with time over the lifetime of an individual LAE, and
between different LAEs, possibly dependent on properties of the host
galaxy such as its mass.

The astrophysical effects are clearly well beyond definitive detailed
modelling given the current understanding of the physical properties
of LAEs and their immediate environments. Several detailed \Lya\
radiative transfer computations have been performed of the emission
line leaving the galaxy and its CGM under varying assumptions. Whilst
the radiative transfer through an optically thick slab results in a
double-horned emission line profile \citep{1973MNRAS.162...43H},
radiative transfer through a wind or accretion flow generally
suppresses either the blue side in a wind \citep{2006A&A...460..397V},
or the red side in an accretion flow \citep{2006ApJ...649...14D,
  2006A&A...460..397V}, producing emission that is redward dominated
in the case of outflow or blueward dominated in the case of
infall. Backscattering off the far-side of the gas relative to the
observer results in peaks displaced by as much as twice the outflow or
infall velocities, possibly achieving offsets as great as $\pm400\kms$
\citep{2006A&A...460..397V}.

The details of the emission line profile are sensitive to the origin
of the emission (central point source or extended emitting region),
the total \HI\ column density, the gas temperature and the expansion
or infall velocity, as well as broadening by any turbulence that may
be present. For emission through an expanding shell, and depending on
the \HI\ column density, for a sufficiently large expansion velocity
the redward lobe splits into two, one peaking at about twice the wind
velocity redward of line centre and the other just shortward of line
centre \citep{2006A&A...460..397V}. The profile also may depend on the
orientation of the LAE. For photons generated within a gaseous disk,
when viewed face-on the profile may be dominated by a single lobe,
either blueward of line centre if the source is at the centre of an
accretion flow, or redward if at the centre of a wind. When viewed
edge-on, however, the profile may split into two lobes around line
centre for a sufficiently high internal \HI\ column density
\citep[eg][]{2012AA...546A.111V}.

In reionization simulations, various approaches have been taken to
model the \Lya\ profiles including the effect of IGM
attenuation. These typically assume complete suppression of any
feature blueward of the systemic \Lya\ wavelength of a LAE, and
partial attenuation of the redward feature by the damping wing from
neutral patches in the IGM
\citep{2006MNRAS.365.1012F,2007MNRAS.381...75M,2008MNRAS.391...63I,
  2011MNRAS.414.2139D}. As we are considering the effects of IGM
attenuation after reionization has completed, such a simpified
treatment is inappropriate. Instead the emission line is attenuated
using the mean IGM transmission $e^{-\tau_\alpha}$ at the cosmological
redshift of the LAE. How much of the profile is attenuated, however,
depends on several factors:\ the centroid of the emission line
compared with the systemic velocity of the LAE, the peculiar velocity
of the LAE, the size of a possible \HII\ region produced by the LAE,
gaseous outflow produced by winds from the hosting galaxy, and
cosmological infall around the LAE. The latter, in particular, can
result in attenuation even when the emission line is redward of the
systemic velocity of a comoving LAE \citep{2011ApJ...728...52L,
  2018MNRAS.479.2564W}. The detailed study of
\citet{2011ApJ...728...52L} shows IGM attenuation on both the blue and
red sides of the local cosmological restframe \Lya\ wavelength,
extending redward by more than $200\kms$ at $z=5.8$ in an early
reionization scenario (with characteristic reionization redshift
$z_\mathrm{ri}=10$) and to over $600\kms$ for late reionization
($z_\mathrm{ri}=6$) (their figures 3 and 4). Typically three-quarters
of the \Lya\ radiation emergent from the galaxy is scattered out of
the line of sight by the large-scale IGM for the early reionization
scenario. The measured \Lya\ luminosity is highly sensitive to galaxy
mass and the profile of the emission line emerging from the galaxy,
whilst the amount further attenuated by the IGM varies widely,
depending on the line of sight (their figure 8).

For simplicity we model the emission line as a single gaussian profile
of width $\sigma_l$.  Although the profile should split into two lobes
for a LAE viewed edge-on, we treat it as unresolved, as it will make
no difference to the total \Lya\ luminosity of a symmetric feature:\
the blueward half of the emission line will be suppressed by roughly
the same factor. We allow the centroid to be displaced relative to the
restframe \Lya\ attenuation edge of the IGM. In the event LAEs
photoionize their surroundings, the displacement is relative to the
near (lower redshift) edge of the \HII\ bubble.

The effect of the IGM attenuation is to alter the intrinsic luminosity $L$ of the \Lya\ emission line, giving the observed value
\begin{equation}
    L_\mathrm{obs}=\frac{L}{1+f}\left(1+fe^{-\tau_\alpha}\right).
    \label{eq:Lobs}
\end{equation}
For a gaussian emission line profile with velocity centroid $v$
relative to the IGM attenuation edge and width $\sigma_l$, the factor
$f$ is given by
\begin{equation}
    f(v) = \frac{1-\mathrm{erf}\left(\frac{
          v}{2^{1/2}\sigma_l}\right)}{1+\mathrm{erf}\left(\frac{
        v}{2^{1/2}\sigma_l}\right)}.
\label{eq:f}
\end{equation}
In the limit $ v \gg\sigma_l$ (eg, wind-dominated emission), $f\rightarrow0$ and
$L_\mathrm{obs}\rightarrow L$, while for $-v\gg\sigma_l$ (eg,
infall-dominated emission), $f\rightarrow\infty$ and $L_\mathrm{obs}\rightarrow
Le^{-\tau_\alpha}$. In the limit of a small velocity offset $\vert v\vert\ll 
\sigma_l$, $f\rightarrow1$ (eg, as for a disk seen edge-on), and
$L_\mathrm{obs}\rightarrow L(1+e^{-\tau_\alpha})/2$.

We show below that only in the presence of substantial IGM attenuation
will UVBG fluctuations have much of an effect on the angular
correlation function of the LAEs. At moderate redshifts ($z\sim3$),
little effect is expected; but the effects may be strong at $z>5$. In
accordance with simulations of the IGM attenuation discussed above,
the IGM attenuation edge may be displaced substantially redward of the
cosmological restframe \Lya\ wavelength at these redshifts. The
characteristic LAE emission line FWHM at $z=5.7$ is $265\pm37\kms$
\citep{2010ApJ...723..869O}, corresponding to
$\sigma_l\simeq110\kms$. The combined displacement redward of the LAE
emission line may be by as much as four times this value, allowing
also for infall. Whilst there is evidence that the \Lya\ emission line
itself is redshifted relative to the systemic velocity of the
\Lya-emitting galaxy \citep[eg][]{2011ApJ...730..136M,
  2018MNRAS.478L..60V}, the observations are largely at $z\sim3-4$,
when winds may be driven by higher rates of star formation in more
massive systems. Many of the systems also show blueward emission,
although at a level consistent with outflows. It is unknown whether the
\Lya\ emission is redshifted compared to the systemic velocity of most
LAEs at $z>5$, but even if so, it is unknown whether the gaseous
surroundings at these redshifts are outflowing, or dominated by cosmic
accretion. If outflows dominate, then no strong contribution of UVBG
fluctuations to the clustering of the LAEs is expected. Observations
suggest the impact of intervening IGM scattering on the
measured \Lya\ profile of LAEs increases with redshift \citep{2021ApJ...908...36H}.

\subsection{Spatial correlations of \Lya\ emitters}
\label{subsec:LAExi}

The effect of fluctuations in the density field and the UVBG on the
number counts of LAEs may be quantified using the LAE luminosity
distribution. This is well fit by a Schechter function
\citep{2020ARA&A..58..617O}. Since the luminosity detection
threshold\footnote{The detection threshold is in practice based on
  equivalent width. We simplify the analysis by formulating the
  detection in terms of a luminosity threshold.} is small compared
with the exponential cut-off in the Schechter function, the total
number of detected systems may be approximated as a power-law
$d{\cal N}/dL = n_*L^\alpha$ for LAE \Lya\ luminosity $L$. These
systems will dominate the spatial correlation signal unless the high
luminosity systems cluster much more strongly than the low. There is
some evidence for a statistical correlation between clustering
strength and LAE luminosity at $z<5$ \citep{2003ApJ...582...60O,
  2005ApJ...635L.117O}, but this may arise from AGN, which are not in
as great abundance among LAEs at higher redshifts
\citep{2008ApJS..176..301O}.

To allow for the line-of-sight variation in the amount of IGM
attenuation, we model the relative offset between the LAE emission line and IGM  
attenuation edge as a gaussian random process with centroid velocity  
offset $v_0$ and dispersion $\sigma_v$. The observed distribution
is then given by
\begin{equation}
\frac{d{\cal N}}{dL_\mathrm{obs}} =
\frac{1}{(2\pi\sigma^2_v)^{1/2}}\int_{-\infty}^{\infty}
dv\,e^{-(v-v_0)^2/2\sigma^2_v}
  \frac{d{\cal N}}{dL}\vert_{L(v\vert L_\mathrm{obs})}\frac{dL}{dL_\mathrm{obs}}(v),
  \label{eq:dndL}
\end{equation}
where $L(v\vert L_\mathrm{obs})$ is the intrinsic \Lya\ luminosity $L$
required to obtain an observed luminosity $L_\mathrm{obs}$ given a
velocity offset $v$, according to Eqs.~(\ref{eq:Lobs}) and
(\ref{eq:f}). The number of systems above a particular observed value
$L_0$ becomes (for $\alpha<-1$),

\begin{equation}
    {\cal N}(L_\mathrm{obs}>L_0) \simeq -\frac{\bar n_*}{\alpha+1}L_0^{\alpha+1},
\end{equation} 
where, using Eq.~(\ref{eq:dndL}),
\begin{equation}
\bar n_* =\frac{n_*}{(2\pi\sigma^2_v)^{1/2}}\int_{-\infty}^\infty
dv\,e^{-(v-v_0)^2/2\sigma^2_v}\left[\frac{1+f(v)}{1+f(v)e^{-\tau_\alpha}}\right]^{\alpha+1}.
\label{eq:nsbar}
\end{equation}
Good fits to the observed distribution of LAE luminosities are
provided by $\alpha=-1.8$ for $z\leq5.7$ and $\alpha=-2.5$ for
$z\geq5.7$ \citep{2020ARA&A..58..617O}.

Assuming $n*$ is proportional to the local cosmological mass density,
the perturbation in the number of detected LAEs, allowing for both
mass density and UVBG fluctuations, is then given by

\begin{equation}
    \frac{\hat\delta{\cal N}}{\cal N}(\mathbf{k},z) = \left[b_L  + f(\Omega_m)\mu_k^2\right]\hat\delta(\mathbf{k},z) +
    (\alpha+1)g(v,\tau_\alpha)\hat\delta\tau_\alpha(\mathbf{k},z),
    \label{eq:dlogN}
  \end{equation} 
where $b_L$ is the cosmological density bias factor for the LAEs and
\begin{equation}
    g(v_0,\tau_\alpha)=
    \frac{\int_{-\infty}^\infty 
      dv\, e^{-\frac{(v-v_0)^2}{2\sigma^2_v}}\left[\frac{1+f(v)}{1+f(v)e^{-\tau_\alpha}}\right]^{\alpha+1}\frac{f(v)e^{-\tau_\alpha}}{1+f(v)e^{-\tau_\alpha}}}{\int_{-\infty}^\infty 
      dv\,
      e^{-\frac{(v-v_0)^2}{2\sigma^2_v}}\left[\frac{1+f(v)}{1+f(v)e^{-\tau_\alpha}}\right]^{\alpha+1}}.
    \label{eq:gv}
\end{equation}
This provides a convenient parametrisation of the influence of UVBG
fluctuations on the LAE correlation function in that $g$ is confined
to the range $0\le g\le 1$. For simplicity, unless stated otherwise,
we take the same width for the velocity offset gaussian as for the
line profile ($\sigma_v=\sigma_l$). This avoids a partial degeneracy
between the offset of the centroid and the width of the
distribution. Although we use the full integration in
Eq.~(\ref{eq:gv}) for all results presented below, we note that at
$z=5.7$ and for $\alpha=-1.8$, $g$ is well-approximated over
$-4\sigma_l< v_0< 0$ by
\begin{equation}
  g(v_0,z=5.7)\simeq\frac{1}{2}\left[1-\tanh\left(\frac{3+v_0/\sigma_l}{2^{1/2}}\right)\right].
    \label{eq:gapprox}
\end{equation}
The additional term $f(\Omega_m)\mu_k^2$ in Eq.~(\ref{eq:dlogN}) accounts for
redshift space distortions \citep{1987MNRAS.227....1K}, where
$f(\Omega_m)$ is the dimensionless linear growth rate of the density
fluctuation growing mode (a flat Universe is assumed) and
$\mu_k=\mathbf{\hat n}\cdot\mathbf{\hat k}$ for line-of-sight
direction $\mathbf{\hat n}$.

The LAE power spectrum, allowing for both density and
UVBG fluctuations and for redshift space distortions, is then

\begin{eqnarray}
  P_L(k,\mu_k,z) &=& V_u\left\langle \frac{\delta\hat {\cal N}^*}{\cal N}({\bf
  k},z)\frac{\delta\hat{\cal N}}{\cal N}({\bf k},z)
               \right\rangle\nonumber\\           
     &=&b_\mathrm{eff}^2V_u \left\langle
         \hat\delta^*(\mathbf{k},z)\hat\delta(\mathbf{k},z)
         \right\rangle \left[1+\frac{f(\Omega_m)}{b_\mathrm{eff}}\mu^2_k\right]^2\nonumber\\
  &+&b_{\alpha,\Gamma}(\alpha+1)g\tau_\alpha b_\mathrm{eff} V_u\Biggl[ \left\langle
                                                   \hat\delta^*_\Gamma(\mathbf{k},z)\hat\delta(\mathbf{k},z)\right\rangle\phantom{\Biggr]}
                                                   \nonumber\\
  &&\phantom{\Biggl[}+ \left\langle \hat\delta_\Gamma(\mathbf{ k},z)
     \hat\delta^*(\mathbf{k},z) \right\rangle
     \Biggr]\left[1+\frac{f(\Omega_m)}{b_\mathrm{eff}}\mu^2_k\right]\nonumber\\
  &+& \left[b_{\alpha,\Gamma}(\alpha+1)g\tau_\alpha\right]^2
    V_u\left\langle\hat\delta^*_\Gamma(\mathbf{k},z)\hat\delta_\Gamma(\mathbf{k},z)
    \right\rangle,
    \label{eq:PkLAE}
\end{eqnarray}
where
\begin{equation}  
b_\mathrm{eff}=b_L+b_{\alpha,\delta}(\alpha+1)g\tau_\alpha,  
\label{eq:beff}
\end{equation}  
and we have introduced the bias factors
$b_{\alpha,\delta}=d\log\tau_\alpha/d\log\rho_b$ and
$b_{\alpha,\Gamma}=d\log\tau_\alpha/d\log\Gamma$, where $\rho_b$ is
the baryon density. The factor $b_\textrm{eff}$ includes the
  radiative transfer selection effect on the LAE sample discussed by
  \citet{2011ApJ...726...38Z} and \citet{2011MNRAS.415.3929W}, as an
  enhancement in the IGM density reduces the detectability of a
  LAE. For $\alpha<-1$, this results in a suppression of the LAE
  clustering signal. Because fluctuations in the UVBG will also affect
  $\tau_\alpha$, additional terms proportional to $b_{\alpha,\Gamma}$
  contribute. These arise from fluctuations in the mean free path of
  Lyman continuum photons and from fluctuations in the spatial density
  of the radiation sources, here assumed to trace the underyling dark
  matter density field. The additional contributions are proportional
  to $\hat\delta^*_\Gamma \hat\delta$ (and its complex
  conjugate). Another contribution arises from the shot noise of the
  sources, which is included in the term proportional to
  $\hat\delta^*_\Gamma\hat\delta_\Gamma$.  \citep[See][for
  details.]{2019MNRAS.482.4777M}

The values for $b_{\alpha,\delta}$ and $b_{\alpha,\Gamma}$ are
computed for the log-normal approximation model for the Ly$\alpha$
forest described in \citet{2020MNRAS.491.4884M}, for which the baryon
fluctuations are the same as the dark matter fluctuations over the
length scales of interest (large compared with the Jeans length of the
IGM). The redshift dependences of the coefficients are well fit over
$2<z<6$ by
$$
b_{\alpha,\delta}\simeq-0.07834(1+z)+1.12334
$$
and
$$
b_{\alpha,\Gamma}\simeq0.07645(1+z)-0.71815.
$$

The computation of the predicted angular correlation function is 
facilitated by decomposing the power spectrum, including redshift space 
distortions, into its Legendre components 
\begin{equation}
P_{\it l}(k,z) = \frac{2{\it 
    l}+1}{2}\int_{-1}^1\,d\mu_k\,P_L(k,\mu_k,z)L_{\it l}(\mu_k), 
\label{eq:Plkz}
\end{equation}
\citep{1998ASSL..231..185H}, where $L_{\it l}(\mu_k)$ is a Legendre 
polynomial of order ${\it l}$. In the linear density approximation 
(and assuming a flat sky) only the ${\it l}=0, 2$ and 4 components are 
non-vanishing. The Legendre components of the redshift-space correlation function 
corresponding to the power spectrum components are given by 
\begin{equation}
\xi_{\it l}(r,z) = \frac{i^{\it 
    l}}{2\pi^2}\int_0^\infty\,dk\,k^2j_{\it l}(kr)P_{\it l}(k,z). 
\label{eq:xirz}
\end{equation}
The angular dependence may be recovered through 
\begin{equation}
\xi(r,\mu,z)=\sum_{i=0}^2 L_{2i}(\mu)\xi_{2i}(r,z), 
\label{eq:ximurz}
\end{equation}
where $\mu=\mathbf{\hat n} \cdot \mathbf{\hat r}$ for two points
separated by $\mathbf{r}$. The angular correlation function at redshift $z$ for objects
separated by a ray with projected separation $b_\perp\simeq D_A\theta$
(in the small angle approximation) relative to the
line of sight from the observer is then
\begin{equation}
    \omega(\theta,\mu)\simeq\sum_{i=0}^{2}\xi_{2i}(D_A[1+z]\theta/[1-\mu^2]^{1/2})L_{2i}(\mu),
    \label{eq:omth}
\end{equation}
where $D_A$ is the angular-diameter distance to redshift $z$ and $\mu$
is the cosine of the angle between the ray connecting the objects and
the line of sight. The redshift factor $[1+z]$ multiplying $D_A$ is
introduced since $k$ and $r$ are assumed to be in comoving units in
Eq.~(\ref{eq:xirz}).

Measurements of the angular correlation function of LAEs are carried
out for systems confined to redshift shells defined by the narrow-band
filter used for measuring the \Lya\ emission line. A shell of width
$\Delta z$ in redshift corresponds to a comoving width
$\Delta {\sc l} = (c/H)\Delta z$.  To compute the angular correlation
function in a shell of width $\Delta z$, the spatial correlations must
be averaged along the line of sight through the shell:
\begin{equation}
    \omega(\theta)\simeq\frac{2}{\Delta{\sc
        l}}\sum_{i=0}^{2}\int_0^{\Delta{\sc l}/2}d{\sc
      l}\,\xi_{2i}\left({\sc l}/\mu\right)L_{2i}(\mu),
    \label{eq:omthsh}
\end{equation}
where $\mu={\sc l}/[{\sc l}^2+(D_A[1+z]\theta)^2]^{1/2}$.

\section{Results}
\label{sec:results}

\begin{figure}
  \scalebox{0.45}{\includegraphics{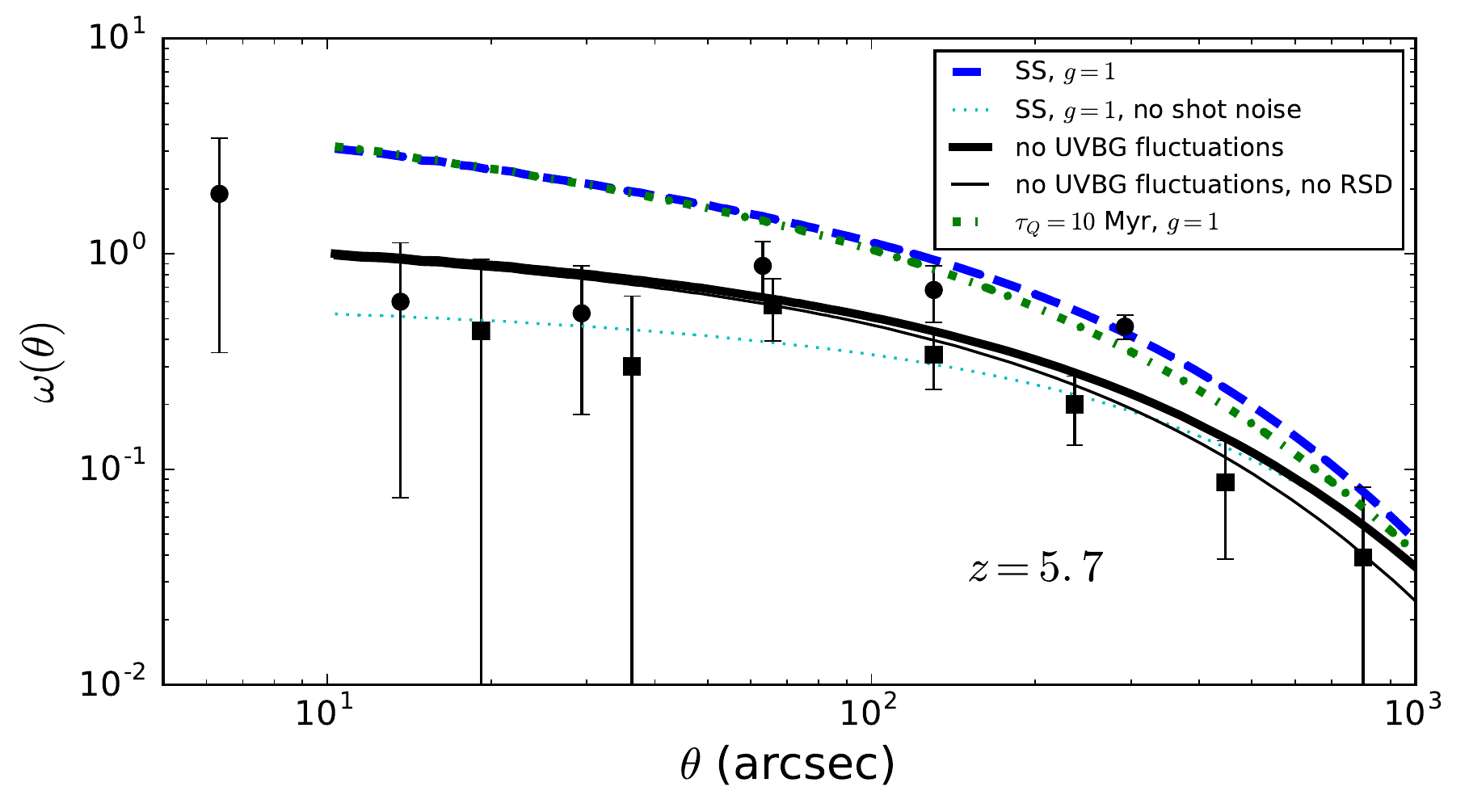}}
  \caption{LAE angular correlation function at z = 5.7 for LAE bias
    factor $b_\mathrm{LAE}=4.1$ and allowing for the maximum effect of
    UVBG fluctuations ($g=1$). Results are shown in the steady-state
    limit with the contribution from source shot noise (long-dashed
    blue curve) and without (dotted cyan curve), and for time varying
    UVBG fluctuations with a QSO lifetime $\tau_Q=10$~Myr (dot-dashed
    green curve). Also shown is the result for no UVBG fluctuations,
    (solid black curves; the lighter curve is without redshift space
    distortions). The data points are taken from
    \citet{2020ARA&A..58..617O}, based on the samples analysed in
    \citet{2010ApJ...723..869O} (circles) and
    \citet{2018PASJ...70S..13O} (squares). (At $z=5.7$, 100~arcsecs
    corresponds to a comoving separation of $2.7h^{-1}$~Mpc.)
  }
  \label{fig:wLAE_SS}
\end{figure}

We first consider the limiting case $g=1$ for the expected angular
correlation function of LAEs, corresponding to
$- v_0\gg\sigma_l$ and $f\rightarrow\infty$, so that the
entire emission line is attenuated, giving the maximal signature of
UVBG fluctuations on the spatial correlations of the LAEs. For $z<5$,
we find the contribution from UVBG fluctuations is negligible (less
than 2\%), but the contribution grows rapidly with redshift as a
result of the rapidly rising intergalactic effective \Lya\ optical
depth.

In Fig.~\ref{fig:wLAE_SS}, we show the effect of UVBG fluctuations
with $g=1$ on the angular correlation function at $z\simeq5.7$,
matching a redshift value for LAEs measured in the SILVERRUSH Subaru
survey \citep{2018PASJ...70S..13O}. We average the spatial
correlations over a redshift shell $\Delta z\simeq0.092$ wide,
corresponding to the width of the narrow-band filter used in the
survey. Two models for the UVBG fluctuations are illustrated, one for
the steady-state limit and a second time-dependent model with a QSO
source lifetime of $\tau_Q=10$~Myr. (The lifetime of the galaxies is
taken to be 100~Myr, but the finite value has a negligible effect on
the UVBG fluctuations because of the high abundance of the galaxies.)
Also shown (black solid curves) is a model with no UVBG fluctuations
using the best estimate LAE bias factor from
\citet{2018PASJ...70S..13O} of $b_L=4.1$. The light curve does not
include redshift space distortions, which enhance the angular
correlations on large angular scales.\footnote{This unintuitive result
  perhaps deserves some explanation. For small angular separations,
  the displacement between most pairs lies nearly along the line of
  sight, so that $\mu\approx1$ and
  $\xi(r,\mu,z)\rightarrow\xi(r,1,z) =
  \xi_0(r,z)+\xi_2(r,z)+\xi_4(r,z)$. For large angular separations,
  the displacements between pairs are mainly orthogonal to the line of
  sight, so that $\mu\approx0$ and
  $\xi(r,\mu,z)\rightarrow\xi(r,0,z)=\xi_0(r,z)-\xi_2(r,z)/2+3\xi_4(r,z)/8$.
  Without redshift space distortions $\xi_2=\xi_4=0$, and
  $\xi=\xi_0$. Whilst allowing for redshift space distortions boosts
  $\xi_0$, the boost is largely cancelled by $\xi_2(<0)$ in $\xi(r,1,z)$
  on the scales of interest. By contrast, $\xi_2$ boosts
  $\xi(r,0,z)$.}

The UVBG fluctuations greatly boost the correlations on angular scales
$\theta<1000$~arcsec. The enhancement arises primarily from the source
shot noise contribution to the UVBG fluctuations. For a QSO lifetime
of $\tau_Q=10$~Myr, the resulting angular correlation function is
slightly smaller than for the steady-state limit. Decreasing the QSO
lifetime further reduces the strength of the correlations on large
angular scales, with the angle beyond which the correlations are
weakened decreasing with decreasing source lifetime (not
shown). Reducing the LAE bias factor $b_L$ diminishes the strength of
the correlations on large angular scales, with little change at small
angles, where the UVBG fluctuation contribution dominates. It is
noteworthy that the UVBG fluctuations excluding the source shot noise
contribution results in a reduction in the strength of the
correlations on scales smaller than a few hundred arcsecs compared to
the case with no UVBG fluctuations (dotted cyan curve). This is because the
effective bias factor $b_\mathrm{eff}$ in Eq.~(\ref{eq:beff}) is
reduced below $b_L$ by the UVBG fluctuations for $\alpha < -1$:\
overdense regions give rise to an excess in IGM attenuation that
suppresses the detection of LAEs.

\begin{figure}
  \scalebox{0.45}{\includegraphics{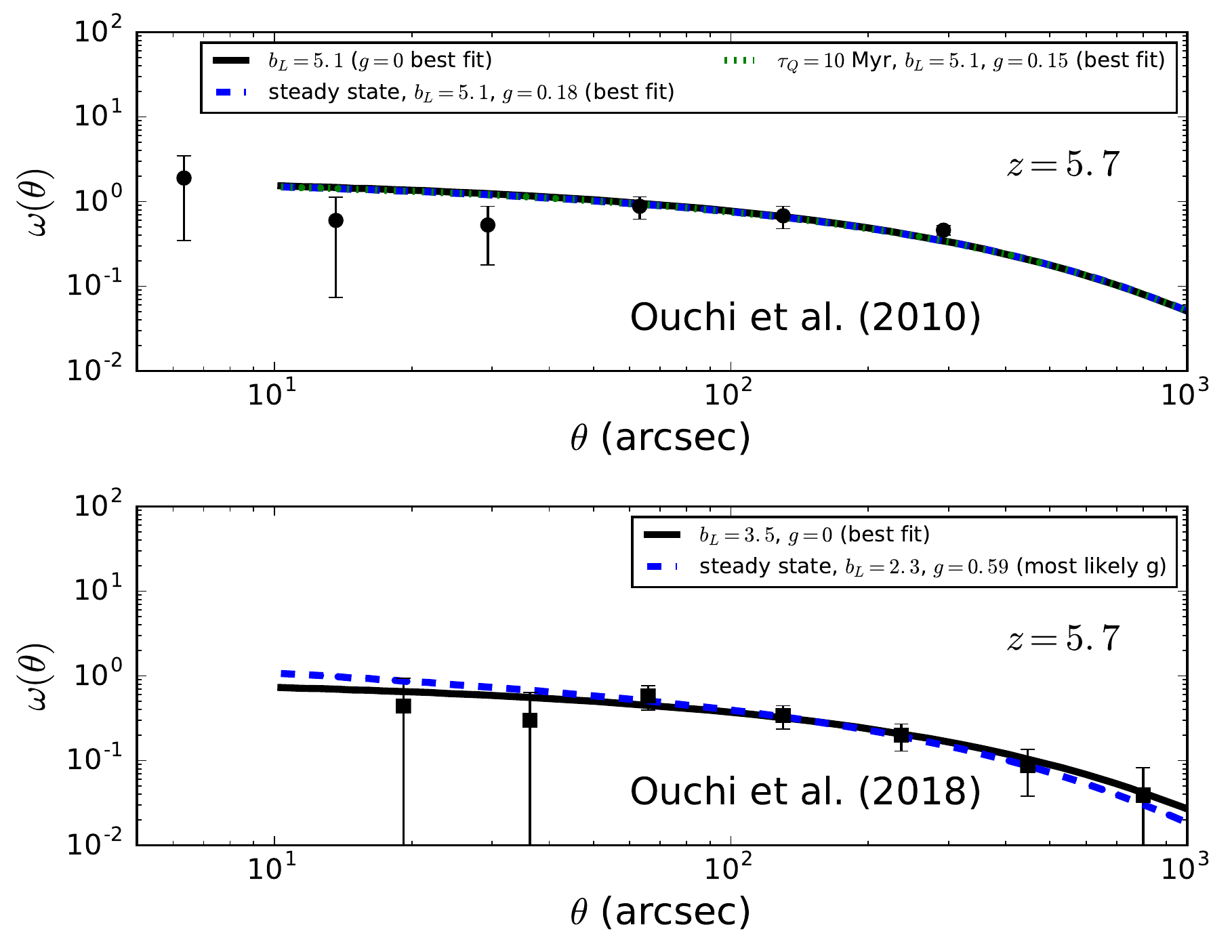}}
  \caption{Best fit models to the measured LAE angular correlation
    functions at $z=5.7$. Upper panel:\ Comparison with the data from
    \citet{2010ApJ...723..869O}. Models are shown without UVBG
    fluctuations ($g=0$; solid black curve), the steady-state model
    (dashed blue curve), and allowing for UVBG fluctuations for a QSO
    lifetime $\tau_Q=10$~Myr (dotted green curve). Lower panel:\
    Comparison with the data from \citet{2018PASJ...70S..13O}. Models
    are shown without UVBG fluctuations ($g=0$; solid black curve) and
    the best-fitting steady-state model for the most likely value of
    $g$, marginalised over $b_L$, corresponding to $b_L=0.23$ and
    $g=0.59$ (dashed blue curve).
  }
  \label{fig:wLAE_bestfits}
\end{figure}

The models are fit to the data by minimising
$\chi^2(b_L,g)=\Sigma_i[\omega_m(\theta_i) -
\omega_d(\theta_i)]^2/\sigma_{d,i}^2$, where $\omega_m$ is the model
prediction for the angular correlation function and $\omega_d$ and
$\sigma_d$ are the measured value and its error
\citep{2018PASJ...70S..13O, 2020ARA&A..58..617O}. Marginalised
probability distributions for $b_L$ or $g$ are computed by integrating
the extraneous variable over the likelihood ${\cal L}=\exp[-\chi^2(b_L,g)/2]$.

\begin{figure}
  \scalebox{0.45}{\includegraphics{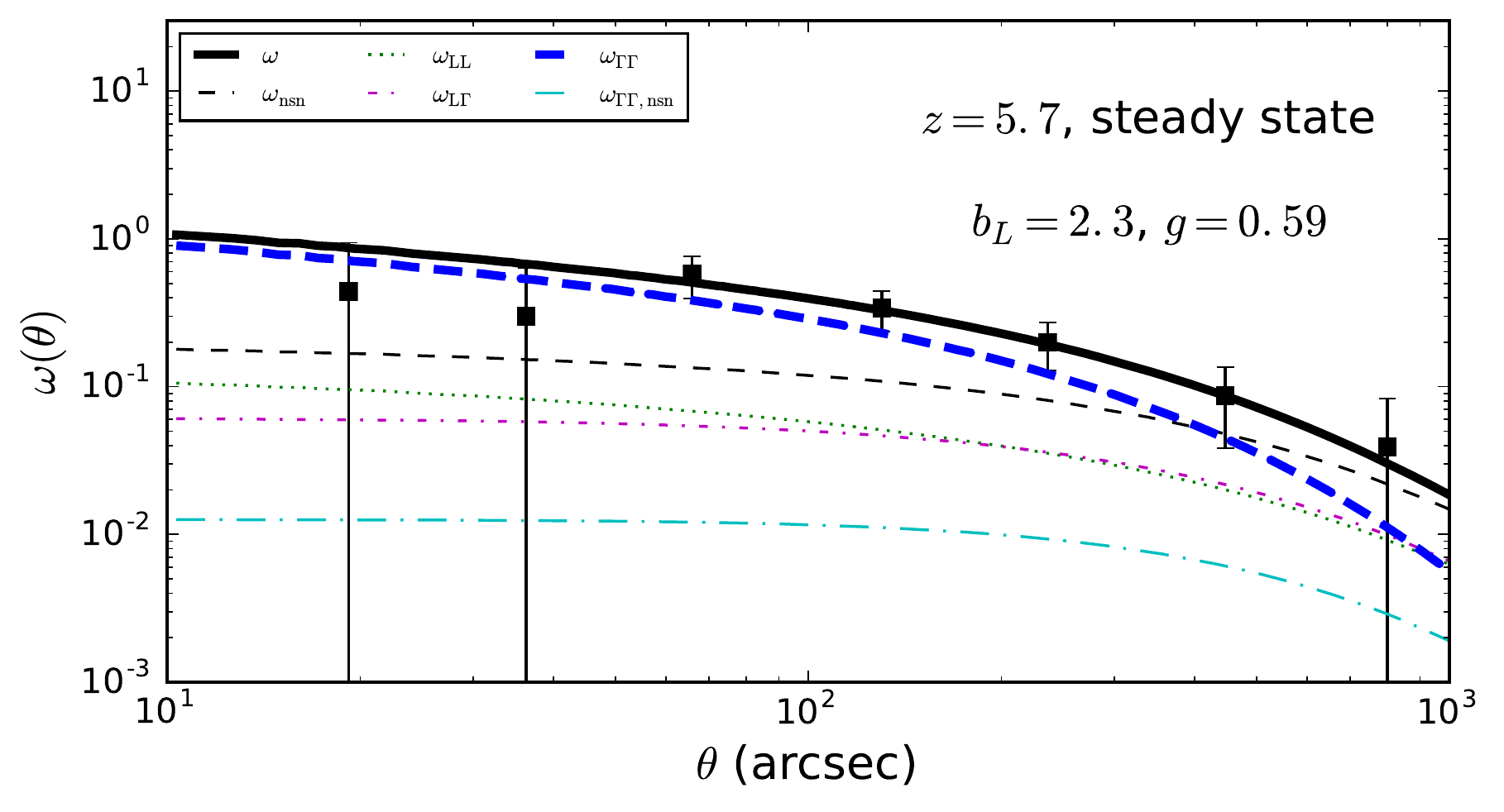}}
  \caption{The break down into component contributions to
    $\omega(\theta)$ for the steady-state model with $b_L=2.3$ and
    $g=0.59$. Shown are the total $\omega$ (solid black curve), the
    total $\omega_\mathrm{nsn}$ excluding the source shot noise
    contribution to the UVBG fluctuations (short-dashed black curve)
    and the separate contributions from the LAE system density
    correlations $\omega_{LL}$ (dotted green curve), the
    cross-correlation $\omega_{L\Gamma}$ between the LAE system
    density fluctuations and the UVBG fluctuations (dot-dashed magenta
    curve), the UVBG fluctuation correlations $\omega_{\Gamma\Gamma}$
    (long-dashed blue curve), and the UVBG fluctuation correlations
    excluding the source shot noise contribution
    $\omega_{\Gamma\Gamma, \mathrm{nsn}}$ (dotted-long-dashed cyan
    curve).
  }
  \label{fig:wLAE_contributions}
\end{figure}

\begin{table}
\centering  
\begin{tabular}{l|c|c|c|r}
  $\tau_Q$ & sample & $b_L$ & $g$ & $\chi^2$ \\
  (Myr) & & & & \\
  \hline\hline  
--  & O18 & $3.5^{+0.3}_{-0.4}\, \left( ^{+0.6}_{-0.8}\right)$
                                              & $0$ & 1.40\\
--  & O10 & $5.1\pm0.3\, \left(^{+0.6}_{-0.7}\right)$
                                              & $0$ & 10.6\\
\hline  
  10   & O18 & $3.4^{+0.4}_{-0.6}\, \left( ^{+0.8}_{-2.0}\right)$
                                              &
                                                $0.40^{+0.13}_{-0.35}\, \left(<0.69\right)$ & 1.40\\
  10 & O10 & $5.1^{+0.3}_{-0.4}\, \left( ^{+0.6}_{-0.8}\right)$
                                              & $<0.29\, 
                                                \left(<0.49\right)$ & 10.1\\
\hline
  $\infty$ & O18 & $3.4^{+0.5}_{-1.0}\, \left( ^{+0.7}_{-2.7}\right)$ &
                                                               $0.59^{+0.09}_{-0.39}\left(<0.70\right)$
    & 1.40\\
 $\infty$  & O10 &$5.0^{+0.4}_{-0.3}\, \left( ^{+0.7}_{-0.8}\right)$ &
                                                             $0.18\pm0.17\left(<0.54\right)$  & 10.6\\
\hline 
\end{tabular}
\caption{Marginalised estimates for LAE bias $b_L$ and
  attenuation factor $g$, without UVBG fluctuations (top section;
  $g=0$), for a QSO lifetime $\tau_Q=$10~Myr (middle section) and the
  steady state model ($\infty$) (bottom section). Estimates based on
  $w(\theta)$ are computed using the data of
  \citet{2010ApJ...723..869O} (O10) and \citet{2018PASJ...70S..13O}
  (O18). Ranges are given for 68\% (95\%) confidence intervals or
  upper limits. The last column gives $\chi^2$ for the best-fitting
  model.
}
\label{tab:bg_fits}
\end{table}

Best fits to the measured correlations at $z=5.7$ (including the
integral correction for the finite survey areas), are shown in
Fig.~\ref{fig:wLAE_bestfits}, using the measured values from
\citet{2010ApJ...723..869O} (O10) and \citet{2018PASJ...70S..13O}
(O18). The point from O10 with lowest angular separation is excluded
since the dark matter density fluctuations become non-linear on this
scale \citep{2018PASJ...70S..13O}. The best-fit models to the O10
result are found to nearly coincide with the best-fitting model with
no UVBG fluctuations ($g=0$). The best-fitting model to the O18 result
has $g=0$ for both the steady state and $\tau_Q=10$~Myr cases. Also
shown is the best-fitting model fixing $g=0.59$, corresponding to the
most likely value for $g$ after marginalising over $b_L$ in the
steady-state model, as given in Table~\ref{tab:bg_fits}. For $g=0.59$,
$\chi^2$ is minimum at $b_L=2.3$. The difference from the overall best
model curves shows that much of the weight of the fits comes from the
data at small angle separations. This is illustrated in
Fig.~\ref{fig:wLAE_contributions} for the model with $b_L=2.3$ and
$g=0.59$. The curve for $\omega_{\Gamma\Gamma}$ shows that the
correlations become increasingly dominated by the UVBG fluctuations on
scales below 400 arcsecs. A comparison between $\omega_{\mathrm{nsn}}$
and the full correlation function $\omega$ shows the fluctuations are
dominated by the contribution to the UVBG fluctuations from the source
shot noise on scales below 100 arcsecs, given by the difference
between $\omega_{\Gamma\Gamma}$ and
$\omega_{\Gamma\Gamma, \mathrm{nsn}}$. The density-dependent radiative
transfer selection effect is subdominant on these scales for this model.

\begin{figure}
\subfloat{\includegraphics[width=0.25\textwidth, height=0.18\textheight]{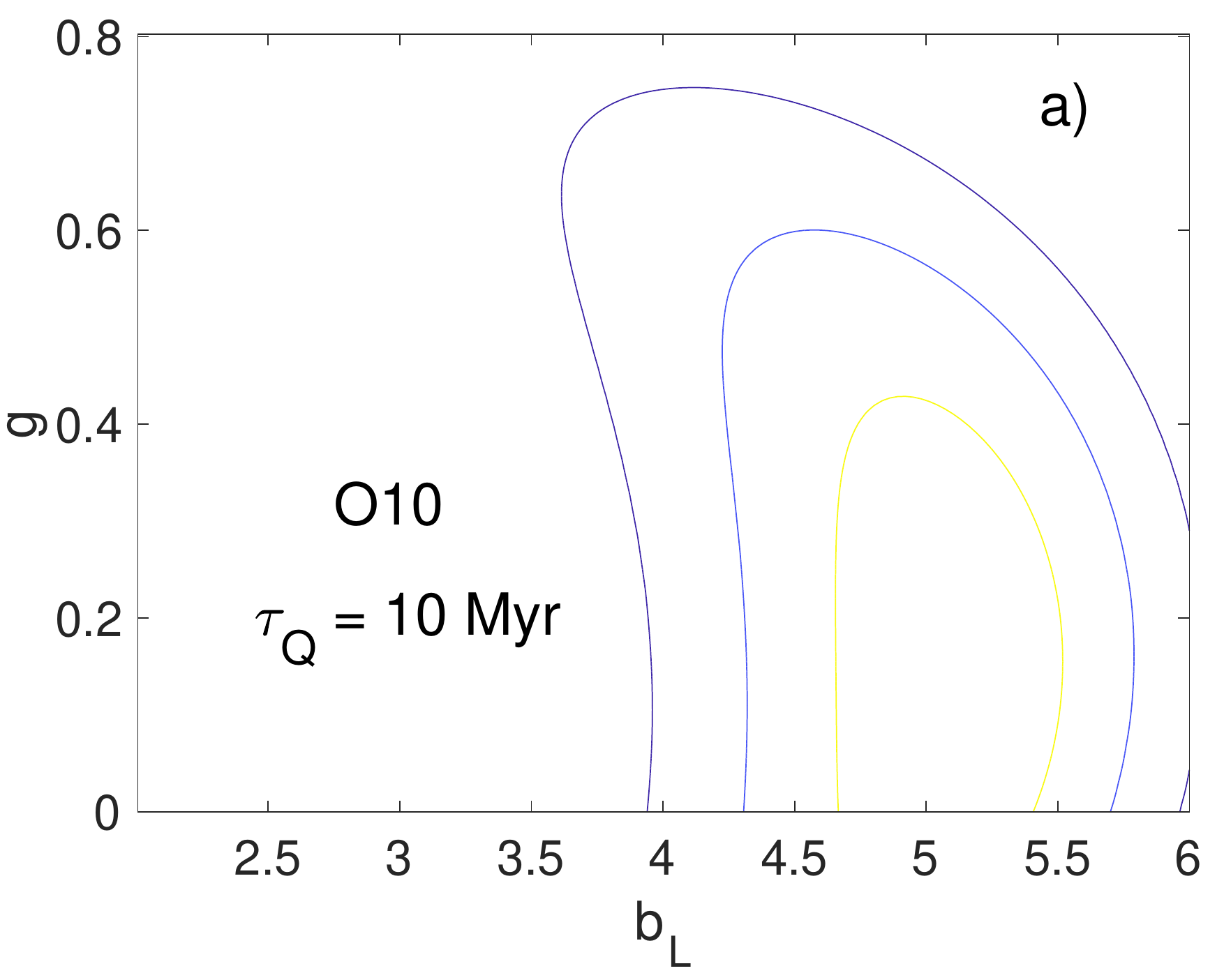}}
\subfloat{\includegraphics[width=0.25\textwidth,height=0.18\textheight]{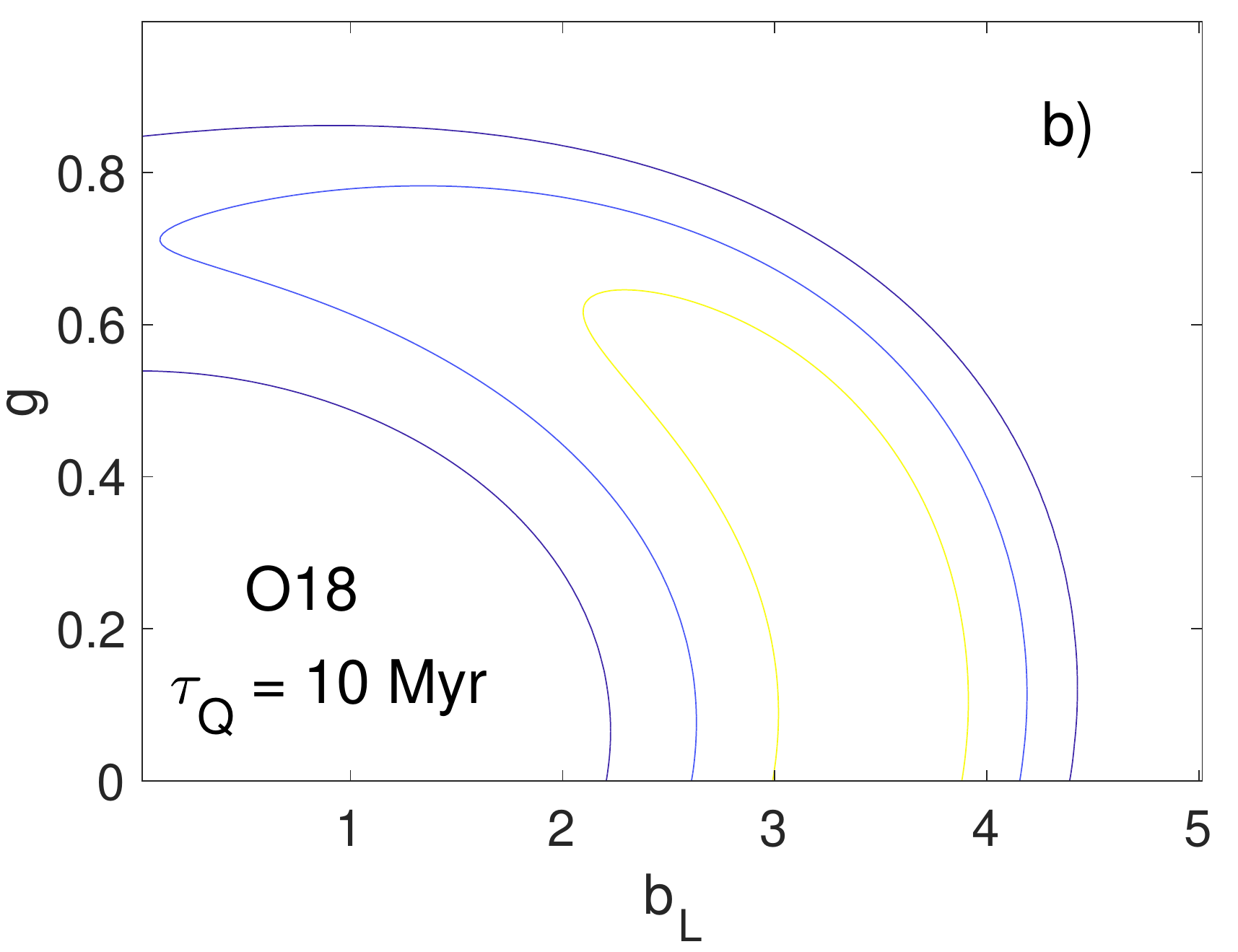}}\\
\noindent  
\subfloat{\includegraphics[width=0.25\textwidth, height=0.18\textheight]{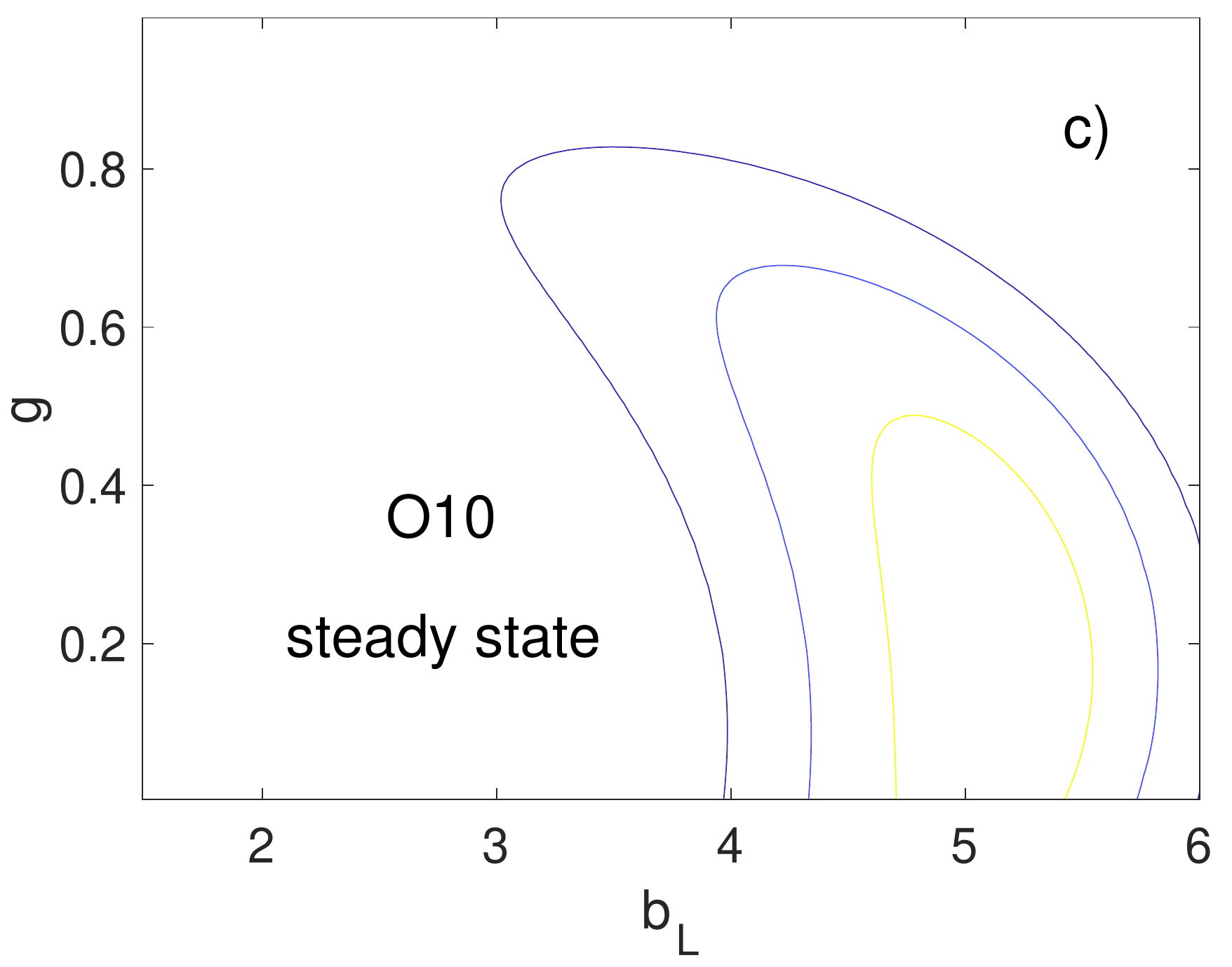}}
\subfloat{\includegraphics[width=0.25\textwidth,height=0.18\textheight]{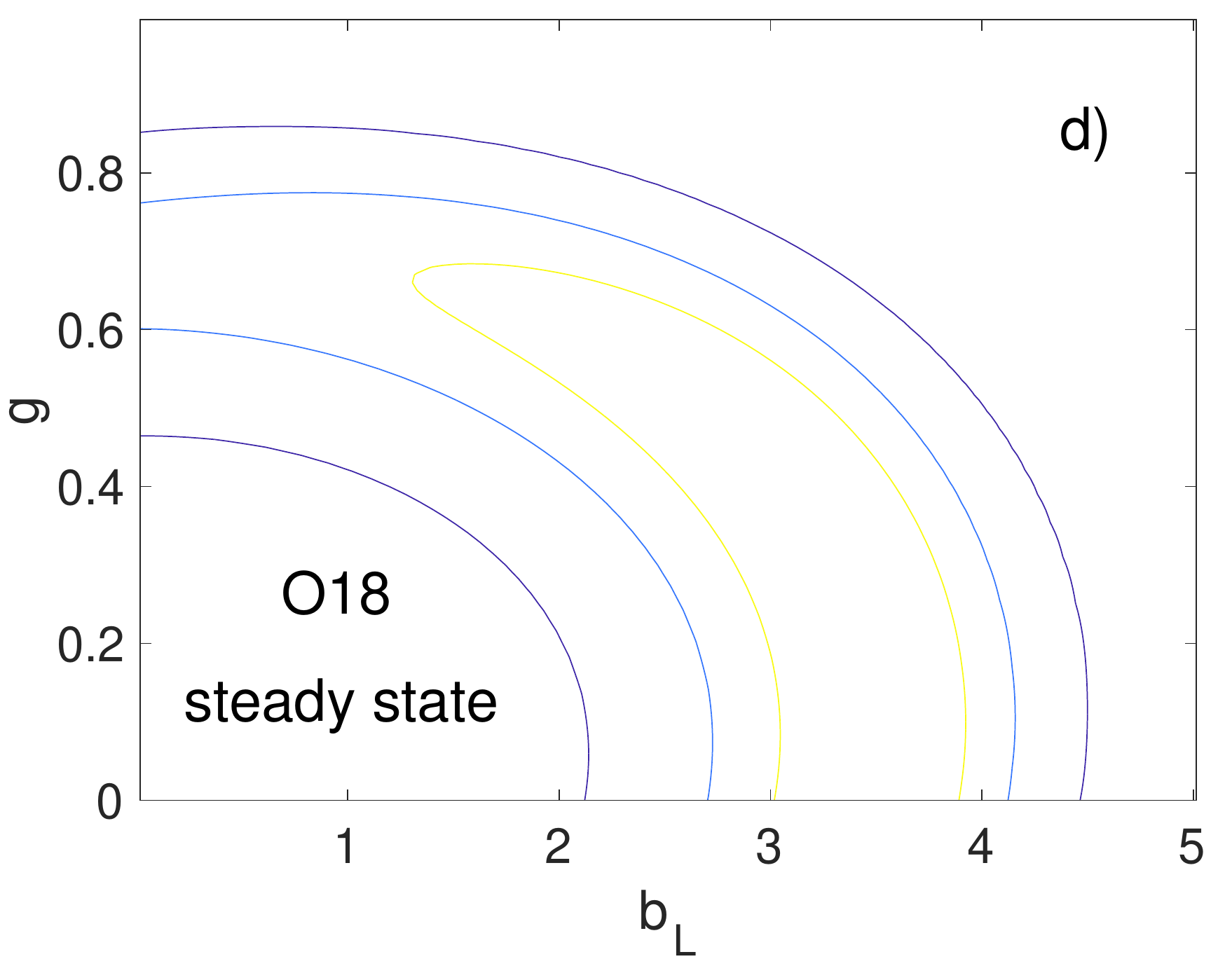}}
\caption{Likelihood contour levels in $b_L$ and $g$ for fits to the 
  measured angular correlation functions at $z=5.7$. Results shown for 
  UVBG fluctuation models with $\tau_Q=10$~Myr to the data from a) O10 
  and b) O18, and for the steady-state limit for the data from c) O10 
  and d) O18. (The contour levels correspond to 1$\sigma$, 2$\sigma$
  and 3$\sigma$ enclosed probabilities.) 
}
\label{fig:wLAE_contours}
\end{figure}

Likelihood contours for the fits are presented in
Fig.~\ref{fig:wLAE_contours}. Allowing for the UVBG fluctuations opens
up solutions with low LAE bias $b_L$ and high $g$, particularly for
the O18 data. For both the O10 and O18 data, low $b_L$ and high $g$
solutions become relatively more preferable for the steady state model
compared with the $\tau_Q=10$~Myr model.

The estimates for the LAE bias $b_L$ (marginalised over $g$) and for
$g$ (marginalised over $b_L$) are displayed in Table~\ref{tab:bg_fits}
for a QSO lifetime $\tau_Q=10$~Myr and in the steady-state limit. A
bias of $b_L\approx3.4$ is found for the O18 data, and $b_L\approx5$
from the smaller O10 sample, with agreement at the 68\% confidence
level regardless of the assumed QSO lifetime, or excluding UVBG
fluctuations ($g=0$). At the 95\% confidence limits, however, the
solutions open up to allowing substantially reduced values for $b_L$
with $g$ as high as $\sim0.5$ for the O10 sample and $\sim0.7$ for the
O18 sample.

Because much of the weight for the increased range in $g$ arises from
the correlations at small angles, and the best-fitting models lie
systematically above the data points at small angles (see
Fig.~\ref{fig:wLAE_bestfits}), we also provide results using an
alternative fitting approach based on allowing for possible
underestimates in the error bars following
\citet{2010arXiv1008.4686H}. For the steady-state model fit to the O10
sample, the model parameter expectation values and 68\% (95\%)
confidence intervals obtained are
$b_L=5.0^{+0.4}_{-0.6}\left(^{+0.5}_{-1.2}\right)$ and
$g=0.32^{+0.24}_{-0.21}\left(^{+0.32}_{-0.27}\right)$, while for the
O18 sample, the expectation values obtained are
$b_L = 3.2^{+0.6}_{-1.3} \left( ^{+0.8}_{-2.4} \right)$ and
$g=0.46\pm 0.28\, \left(^{+0.32}_{-0.41} \right)$. These are
comparable to the values found from the more direct maximum likelihood
approach above, although with somewhat extended error bars allowing
for a larger contribution from UVBG fluctuations. See the Appendix for
further details.

\section{Discussion}
\label{sec:discussion}

Whilst UVBG fluctuations are too small to affect the clustering of
LAEs at redshifts $z<5$, allowing for UVBG fluctuations at higher
redshifts is found to open up the LAE bias factors inferred from LAE
clustering to a broader range, especially at the 95\% confidence
interval (CI) level. Assuming no UVBG fluctuations ($g=0$) at $z=5.7$,
we find $b_L\simeq 3.5\pm0.3$ ($^{+0.6}_{-0.8}$, 95\% CI) using the
O18 LAE sample, and $b_L\simeq5.1\pm0.3$ ($^{+0.6}_{-0.7}$, 95\% CI)
using the O10 sample. Allowing for UVBG fluctuations broadens the
error ranges to $b_L\simeq 3.4^{+0.5}_{-0.8}$ ($^{+0.7}_{-2.3}$, 95\%
CI) for the O18 sample, and $b_L\simeq5.0\pm0.3$ ($\pm0.7$, 95\% CI)
for the O10 sample. At the 95\% confidence level, the angular
correlations at angular separations smaller than 400 arcsecs may be
dominated by UVBG fluctuations, allowing for solutions with much
reduced LAE bias factors, especially for the larger O18 sample.

The evolution in the bias factors has been used to infer the nature of
the haloes in which LAE systems reside. For a fixed halo mass, the
bias factor increases rapidly with redshift. On this interpretation,
the evolution in bias factors suggests LAEs occupy haloes in the mass
range $10^{10}-10^{12}\,M_\odot$ at $z>2$ and may be the progenitors
of present-day massive elliptical galaxies
\citep{2020ARA&A..58..617O}. There is, however, considerable scatter
of unclear origin in the redshift trend of the bias factors. Whilst
contamination by randomly distributed sources would reduce the
inferred bias factors, it has also been suggested that the small
volumes of some of the surveys may result in a large scatter from
cosmic variance, particularly if a protocluster is in the field
\citep{2010ApJ...723..869O, 2018PASJ...70....4K}. A protocluster
containing luminous photoionizing sources may also increase the \Lya\
transmission of nearby LAEs, enhancing their clustering. For the O18
samples at $z=5.7$, the findings here suggest low bias factors
$b_L\sim2$ may be consistent with the angular correlation strengths in
the presence of a large UVBG fluctuation contribution. In this case,
the evolution of the bias factor of LAEs is consistent with a constant
halo mass of about $10^{10.5}\msun$ \citep[see fig.16 of][]{2020ARA&A..58..617O}.

The statistical limits on $g$ are very broad. The 68\% confidence
levels nearly include $g=0$. For the O18 sample, the overall
best-fitting models are for $g=0$. Nonetheless, values as high as
$g<0.7$ are acceptable at the 95\% confidence level for the O18
sample, and $g<0.5$ for the O10 sample.

The upper limits on $g$ of 0.5--0.7 correspond to upper limits on the
blueward velocity offset of $3-3.4\sigma_l$, where $\sigma_l$ is the
line width of the emission line, for the LAE emission compared with
the \Lya\ attenuation edge of the surrounding gas. This corresponds to
the higher values found in the simulations of
\citet{2011ApJ...728...52L} for $\sigma_l\approx100-150\kms$ in an
early reionization model ($z_\mathrm{ri}=10$), but are typical for
late cosmic reionization ($z_\mathrm{ri}=6$). Values of
$g\simeq0.05-0.1$ correspond to blueward displacements of the LAE
emission line of $0.9-1.5\sigma_l$, typical for early reionization.

\begin{figure}
  \scalebox{0.45}{\includegraphics{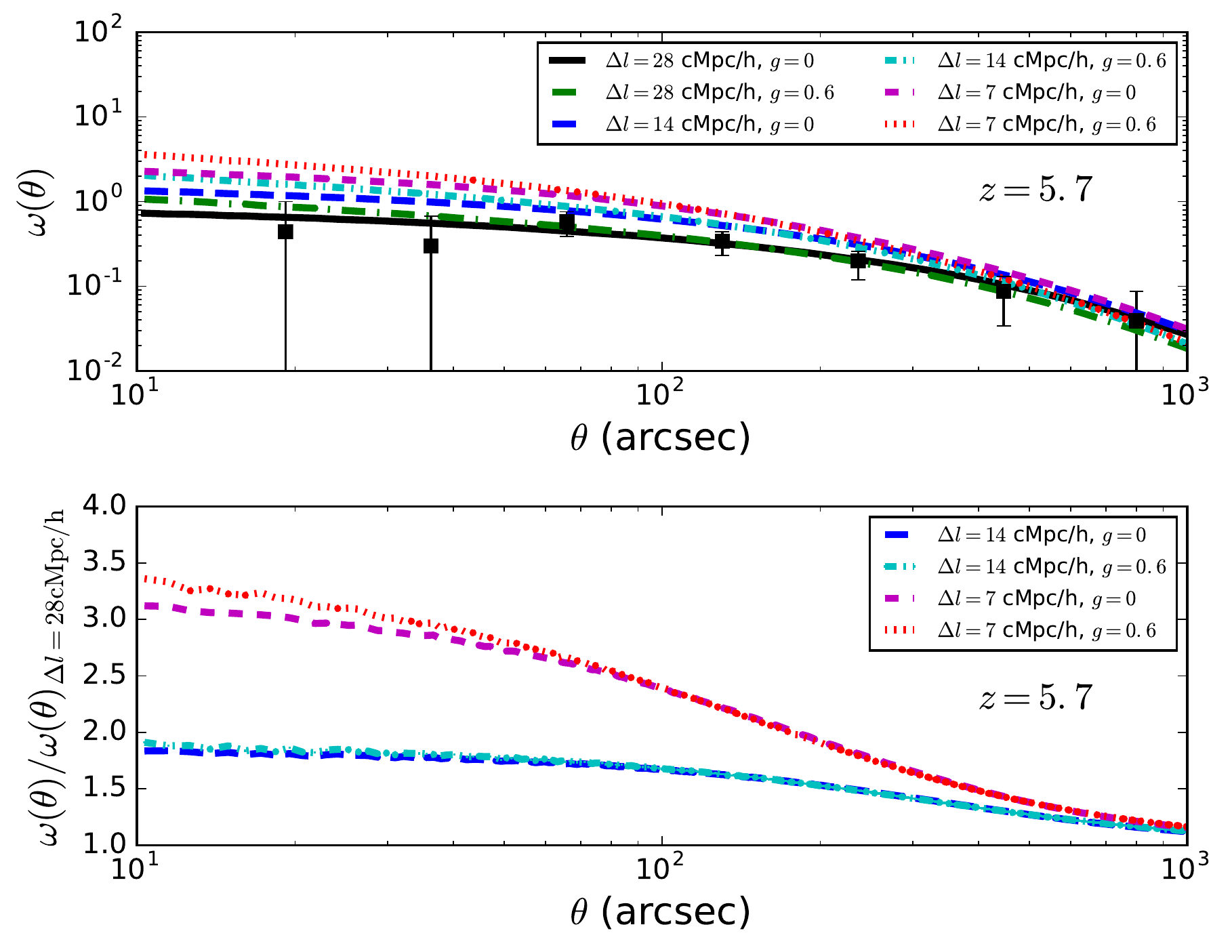}}
  \caption{Dependence of the angular correlation function on width of  
    the redshift shell. Upper panel:\ Angular correlation function  
    without ($b=3.5$, $g=0$) and with ($b=2.3$, $g=0.6$) UVBG  
    fluctuations in the steady-state limit, for shell widths of 28, 14  
    and 7~$h^{-1}\,$Mpc (comoving). The data points are from  
    \citet{2018PASJ...70S..13O}. Lower panel:\ The ratio of the  
    angular correlation functions for shell widths of 14 and 7  
    ~$h^{-1}\,$Mpc, to the correlation functions for 28~$h^{-1}\,$Mpc.  
  }
  \label{fig:wLAE_varied_sigr}
\end{figure}
 
The angular correlation function on small scales is suppressed by the
finite width of the redshift shell imposed by the narrow-band filter
centred on the \Lya\ wavelength for a given redshift. The width of the
redshift shell at $z=5.7$ corresponds to an averaging of the spatial
correlation function up to separations of nearly 30$h^{-1}$~Mpc
(comoving). As a possible means of distingishing between a correlation
function dominated by the (biassed) underlying dark matter density
fluctuation field and by the UVBG fluctuations, we show in
Fig.~\ref{fig:wLAE_varied_sigr} the effect of choosing narrower
filters corresponding to a half and a quarter of the actual filter
width. Using the best-fit models without and with UVBG fluctuations
for the steady state model, the resulting trends are virtually
identical for the O10 data and exactly identical for the O18 data
(since the best-fit model allowing UVBG fluctuations has $g=0$). For
this reason, we show instead the trend for the best-fit model for the
most likely value of $g$, corresponding to $b_L=2.3$ and $g=0.59$, for
the O18 data. For this model, UVBG fluctuations dominate the angular
correlations on angular scales smaller than 400 arcsecs, with the shot
noise contribution dominating on scales smaller than 100
arcsecs. Whilst reducing the thickness of the redshift shell increases
the strength of the correlations (upper panel), the trends are nearly
identical whether or not UVBG fluctuations dominate the correlations
(lower panel). A small difference may be distinuished for the
quarter-width filter case, but this would require measuring the
correlations on these scales to better than 10\% accuracy.

On the other hand, the result shows that the rise in the strength of
the angular correlations is robust when sub-samples analysed are
confined to increasingly narrow redshift ranges. This provides a test
of the possible contribution of contaminating sources to the angular
correlations. Although spectroscopic follow-up suggests a
contamination rate of only $\sim8$\% by foreground objects
\citep{2018PASJ...70S..13O}, if they cluster more strongly than the
LAE systems, they would alter the expected scaling of the angular
correlation function with filter width.\footnote{If the foreground
  objects are weakly or un-clustered, however, they only dilute the
  measured correlations and the scaling would remain unaffected.}

\begin{figure}
\subfloat{\includegraphics[width=0.25\textwidth, height=0.18\textheight]{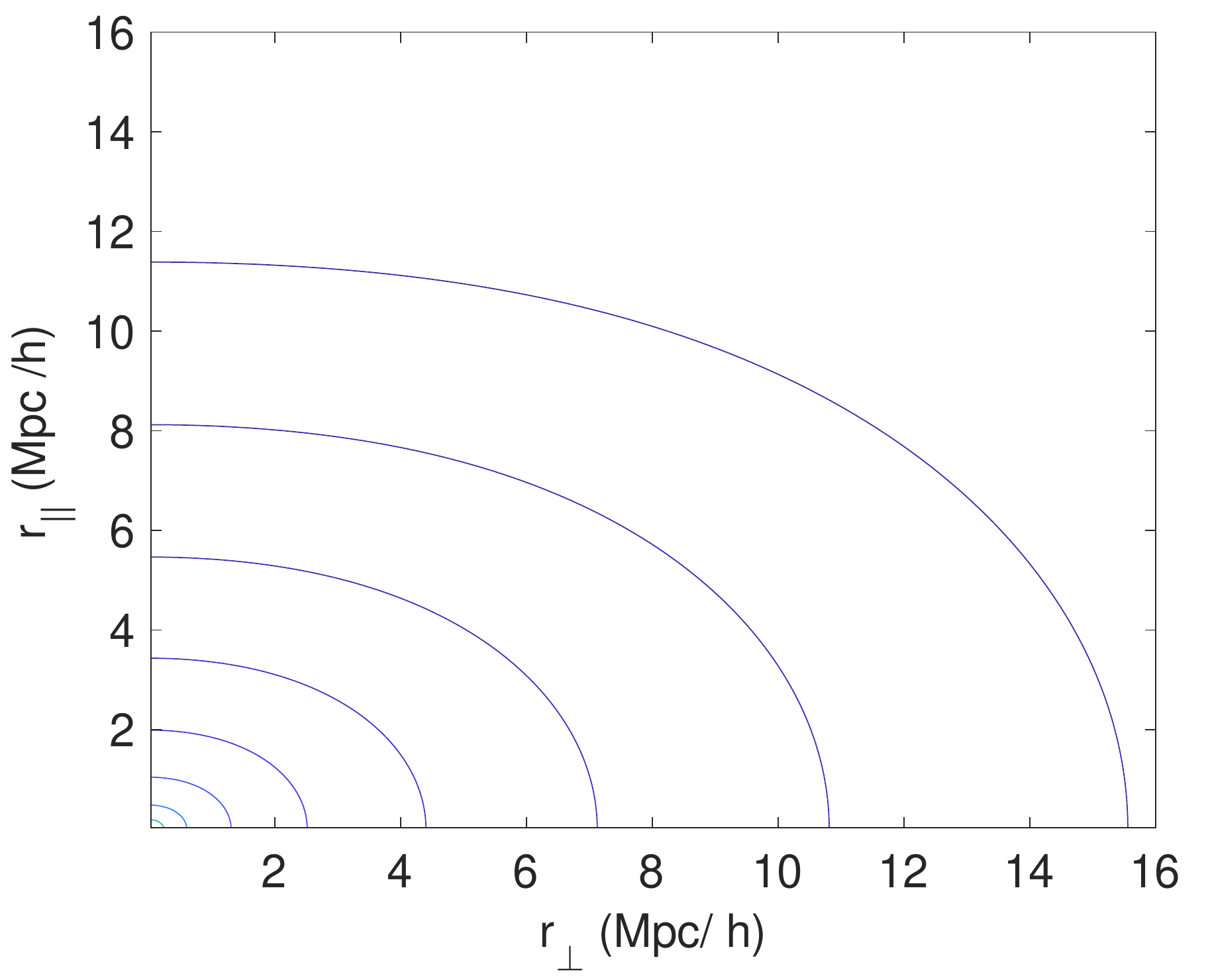}}
\subfloat{\includegraphics[width=0.25\textwidth,height=0.18\textheight]{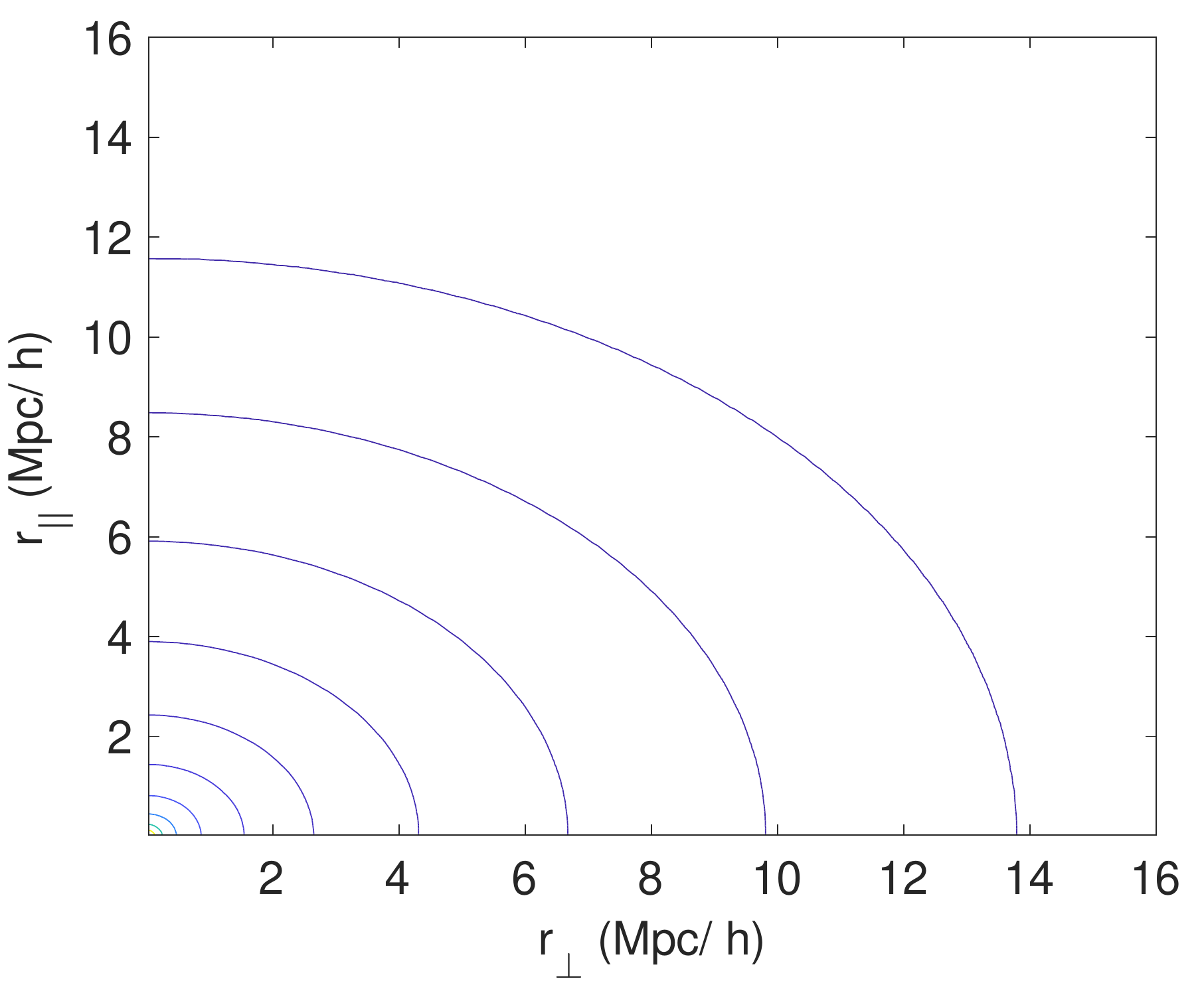}}\
\caption{Contour levels of the spatial correlation function  
  $\xi(r_{||},r_\perp)$, as a function of the separations $r_{||}$
  along the line of sight and $r_\perp$ transverse to the line of  
  sight (in comoving \textrm{Mpc/ h}). Left panel:\ The best fit model  
  to the O18 data for no UVBG fluctuations ($b_L=3.5$). Right panel:\
  The steady-state UVBG fluctuation model with $b_L=2.3$ and  
  $g=0.59$. (The outermost contour level is 0.1, and the levels
  increase inward by multiplicative steps of 1.89.)
}
\label{fig:xilLAE_contours}
\end{figure}

The availability of a large spectroscopic sample would offer the
possiblity of measuring the spatial correlation function
$\xi(r_{||},r_\perp)$ as a function of line-of-sight $r_{||}$ and
transverse $r_\perp$ separations. Fig.~\ref{fig:xilLAE_contours} shows
$\xi(r_{||},r_\perp)$ for the best-fit model without UVBG fluctuations
to the O18 data, with $b_L=3.5$ (left panel) and the steady-state UVBG
model with $b_L=2.3$ and $g=0.59$ (right panel). The angular
correlation functions (Fig.~\ref{fig:wLAE_bestfits}), are nearly
identical. The UVBG fluctuations tend to isotropise
$\xi(r_{||},r_\perp)$ compared with the expectation for a biassed halo
model. Distinguishing the two models, however, would require high
precision redshift determinations.

\section{Conclusions}
\label{sec:conclusions}

We summarise our principal conclusions:

1.\ Allowing for UVBG fluctuations opens up a broader range of
acceptable models for the bias factor of LAEs. For the expected range
in redshifted foreground IGM \Lya\ photon attenuation as may arise
from cosmic infall around the LAE systems in a reionized IGM, UVBG
fluctuations may non-negligibly contribute to the clustering of LAEs
at $z>5$ as measured by the angular correlation function.

2.\ Comparison with the measured angular correlation function of LAEs
in the SILVERRUSH Subaru survey at $z\simeq5.7$ shows that, whilst the
clustering is statistically consistent with no contribution from UVBG
fluctuations, the measured clustering also supports models with a
substantial contribution from UVBG fluctuations. These include models
in which the UVBG fluctuations dominate the clustering signal over
angular separations smaller than a few hundred arcseconds,
corresponding to projected separations smaller than $15h^{-1}$~Mpc
(comoving). On scales smaller than 100~arcsecs (about $3h^{-1}$~Mpc),
the shot noise contribution to the UVBG fluctuations dominates the
correlations in these models. The scales over which shot noise
dominates is smaller for QSO lifetimes shorter than 10~Myr.

3.\ It may be possible to distinguish models with and without a
substantial contribution from UVBG fluctuations to the clustering of
LAEs if the angular correlation function may be measured to 10\%
accuracy on angular scales 10--100~arcsec at $z\simeq5.7$. Another
approach may be to measure the spatial correlations as a function of
line-of-sight and transverse separations, as the UVBG fluctuations
tend to isotropise the correlations.

There is much room for improving the constraint on the contribution of
UVBG fluctuations to the clustering of LAE systems at high
redshift. The estimate provided here made several simplifying
assumptions concerning the luminosity distribution of the LAEs, their
emission line widths and the velocity offset of the emission region
relative to the \Lya\ attenuation edge of the surrounding gas. Our
main conclusion is that UVBG fluctuations are sufficiently large that
they may comprise a significant contribution to the overall LAE
clustering strength at $z>5$ even after the IGM has been
reionized. UVBG fluctuations may continue to contribute to the
clustering of LAEs into the reionization epoch as well, requiring an
extension to models of the role of reionization on the LAE correlation
function. More precise measurements of the LAE correlation function
would better constrain the UVBG fluctuation contribution both during
and after reionization completes, which in turn may help clarify the
properties of the gaseous environment of LAE systems, their feedback
on it, and of the nature of the LAE systems themselves.

\section*{Acknowledgements}

We thank Joanne Cohn and Martin White for helpful discussions, and the
anonymous referee for comments that improved the presentation. TS
gratefully acknowledges support from the UK Science and Technology
Facilities Council, Consolidated Grant ST/R000972/1.

This research made use of the python packages \texttt{emcee}
\citep{emcee} and \texttt{corner.py} \citep{Foreman-Mackey2016}.


\section*{Data Availability}
All data used in this paper are publicly available from the cited references.

\bibliographystyle{mnras}
\bibliography{ms}

\appendix
\section{Alternative error estimator}

Following \citet{2010arXiv1008.4686H}, we define the likelihood as

\begin{equation}
    \ln p(\{\omega_{d,n}\}|\{\theta_n\},b_L,g, f) = -\frac{1}{2} \sum_n \left[
      \frac{(\omega_{d, n}-\omega_{m,n})^2}{s_n^2}+ \ln(2\pi s_n^2)\right]  \textrm{,}
\end{equation}
where
\begin{equation}
    s_n^2 = \sigma^2_n + f^2\omega_{m,n}
\end{equation}
for model correlation $\omega_{m,n}$ at angle $\theta_n$, measured
correlations $\omega_{d,n}$ with error $\sigma_n$, and $f$ is a factor
giving the fractional variance added in the noise model. Due to the
dynamical range of the data, adding a single value to the variance
would overestimate the error on the data points with small correlation
strengths. Instead the variance at any give point is increased by a
fraction of the correlation strength at the same point as predicted by
the model. We use a uniform prior on the logarithm of $f$ instead of
$f$ itself to force $f$ to be non-negative.

\begin{figure}
  \scalebox{0.45}{\includegraphics{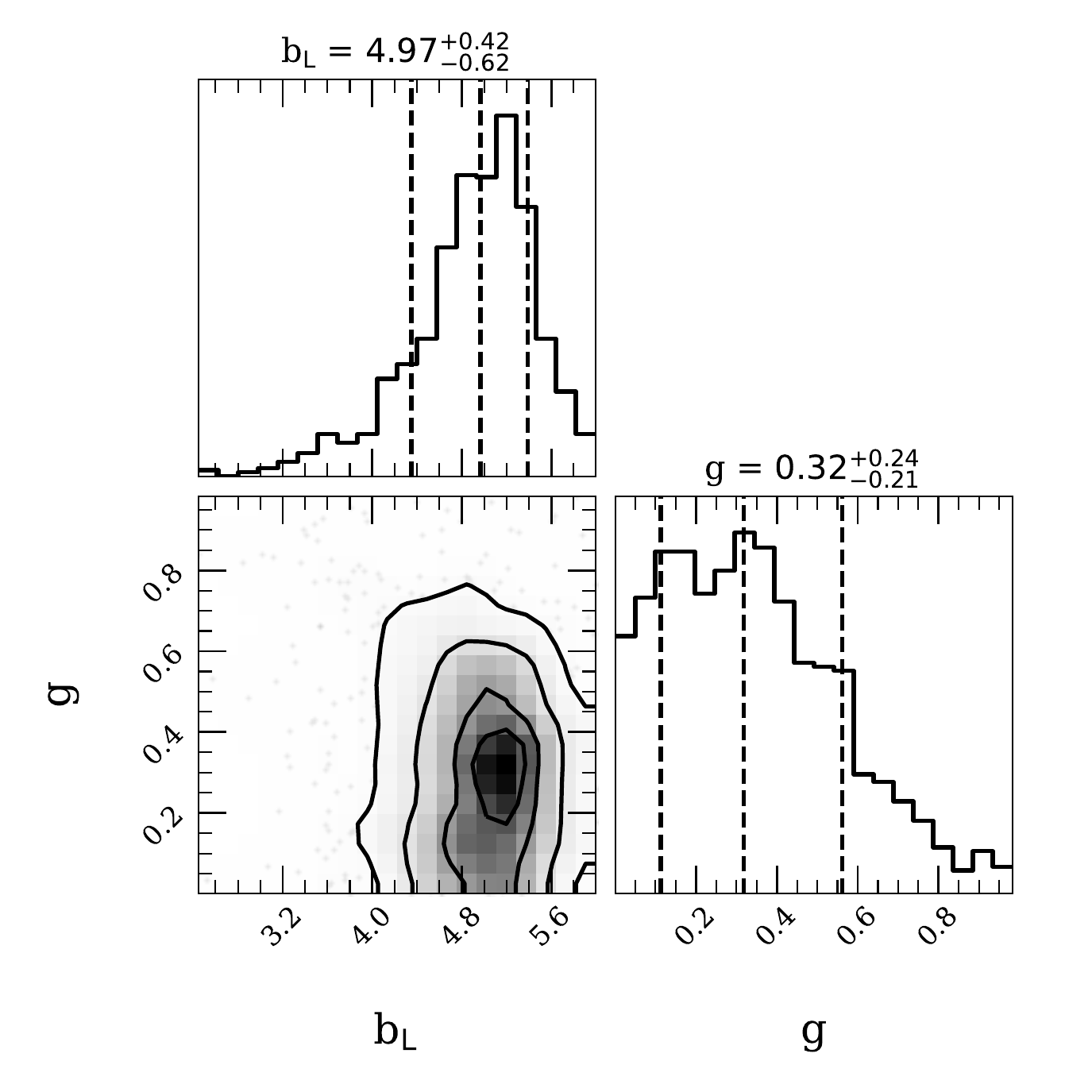}}
  \caption{MCMC parameters fits for the O10 sample. Lower left panel:\
    Likelihood function for $b_L$ and $g$, marginalised over
    $f$. Upper panel:\ Probability density for $b_L$, marginalised
    over $g$ and $f$. Lower right panel:\ Probability density for $g$,
    marginalised over $b_L$ and $f$. The dashed lines in the
    probability distribution plots indicate the expectation values and
    68\% confidence intervals.
  }
  \label{fig:mcmc_circles}
\end{figure}

\begin{figure}
  \scalebox{0.45}{\includegraphics{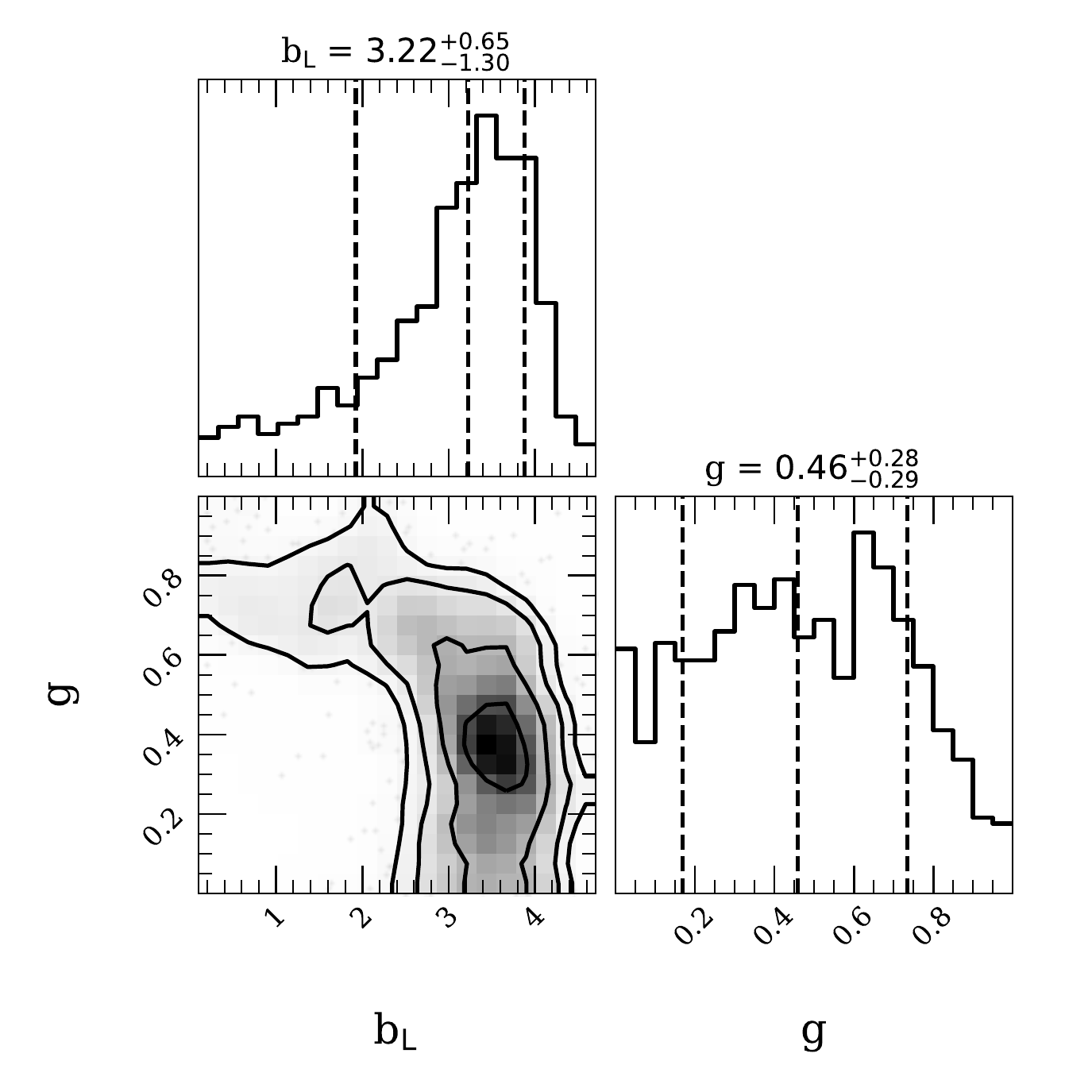}}
  \caption{As for Fig.~\ref{fig:mcmc_circles}, but for the O18 sample.
      }
  \label{fig:mcmc_squares}
\end{figure}

We compute the marginalised expected values for $b_L$ and $g$ using
the Markov Chain Monte Carlo (MCMC) python module
\texttt{emcee}\footnote{\textrm{
    https://emcee.readthedocs.io/en/develop/ }}. The
resulting joint contour plots for $b_L$ and $g$, marginalised over
$f$, and the marginalised distributions for $b_L$ and $g$ are shown in
Figs.~\ref{fig:mcmc_circles} and \ref{fig:mcmc_squares} for the O10
and O18 samples, respectively. The plots were produced using the python module 
\texttt{corner.py}\footnote{\textrm{
    https://github.com/dfm/corner.py }}.

\label{lastpage}
\end{document}